\documentclass[10pt]{article}
\usepackage[utf8]{inputenc}
\usepackage{graphicx}
\usepackage[utf8]{inputenc}
\usepackage{amsmath}
\usepackage{amsfonts}
\usepackage{amssymb}
\usepackage{graphicx}

\usepackage{epstopdf}
\AppendGraphicsExtensions{.tif}

\usepackage{setspace}
\usepackage{authblk}
\usepackage{caption}
\usepackage{float}
\usepackage{cite}
\usepackage[left=2.0cm,right=2.0cm,top=2cm,bottom=2cm]{geometry}
\usepackage{color}
\usepackage{color,soul}
\usepackage{multicol}
\usepackage{subcaption}

\title{\textbf{Turbulence Impacted Wavefront Corrections Using Beam Modulation Technique}}

\author[1,*]{Shouvik Sadhukhan}

\author[2]{C. S. Narayanamurthy}

\affil[1, 2]{\small{Applied and Adaptive Optics Laboratory, Department of Physics, Indian Institute of Space Science and Technology(IIST), P.O: Valiamala, Trivandrum - 695547, State: Kerala; India}}
\affil[1]{\small{Email: shouvikphysics1996@gmail.com}}
\affil[2]{\small{Email: naamu.s@gmail.com}}
\affil[*]{\small{Corresponding Author Email: shouvikphysics1996@gmail.com}}

\begin{document}
\maketitle

\begin{abstract}
New experimental technique have been proposed to discuss the turbulence impact reduction using beam shaping technique. In first phase of experiments, turbulence Impacted Vortex beam shaping technique has been introduced for a propagating beam which is effected by Kolmogorov type lab based turbulence simulator using Pseudo Random Phase Plate (PRPP). The new technique is executed with a spatially filtered collimated Gaussian beam that is propagated through computer controlled rotating PRPP to introduce the lab based turbulence on the beam profile. The turbulence impacted beam is then propagated through Vortex Phase Plate (VPP) to shape that beam into different topological charged Laguerre Gaussian modes and further collimated using two lens based configuration. We have measured the scintillation index of the spatial intensity profile collected from CCD. Comparative studies have been done by placing the image plane in five different positions on the propagation path of the beam so that we can detect Collimated, diverging and converging outputs. Four topological charges of LG beams and Gaussian beam have been used to envisage results.\\

\end{abstract}


\section{Introduction}
Atmospheric turbulence, influenced by factors like temperature and wind, affects the imaging quality of ground-based telescopes and free space optical communication. It creates fluctuations in the refractive index, causing wavefront aberrations in light beams. Two types of turbulence, Kolmogorov and non-Kolmogorov, impact the atmosphere differently. Kolmogorov's theory describes turbulence as homogeneous and isotropic, affecting wavefront aberrations like tilt, focus, and astigmatism. This turbulence leads to beam spreading and wandering, affecting the clarity of images, especially in astronomy. Adaptive optics technology helps counter these effects by dynamically adjusting optical elements, enhancing image quality and enabling better observations of celestial objects. Understanding and mitigating atmospheric turbulence is crucial for improving ground-based astronomy and optical communication in challenging conditions.\cite{1,2,3,4,5,6,7,8,9,10}\par

\begin{figure}[ht]
\centering
\begin{minipage}[b]{1.0\textwidth}
    \includegraphics[width=\textwidth]{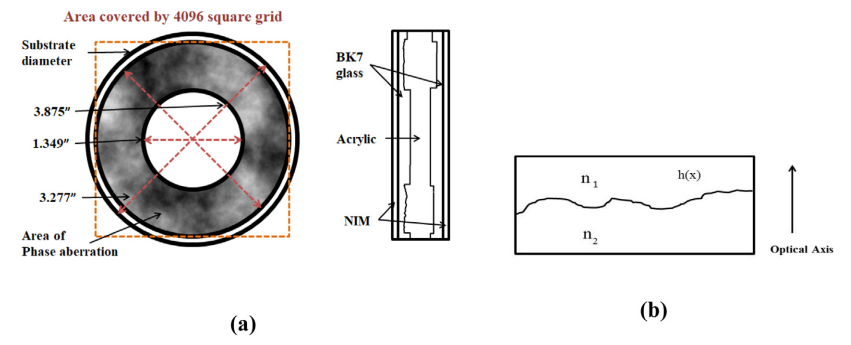}
    \caption{Pictorial representations Pseudo Random Phase Plate}
    \label{Fig: 1}
\end{minipage}
\end{figure}


Optical vortex beams, unlike regular Gaussian beams, have a unique feature known as a singular phase, causing them to have no intensity along their central axis. This property makes them useful for tasks like free space optical communication and optical trapping. Vortex beams carry orbital angular momentum, and when moving through turbulent air, they're less affected than Gaussian beams. Their stability, especially in the face of phase changes, is a big advantage. Some studies, including simulations by researchers like Greg Gbur and Robert K Tyson\cite{13}, show that these vortex beams maintain their integrity even in turbulent conditions. This quality is essential for free space optical communication, where atmospheric turbulence can be a challenge. Even though most studies are simulation-based, they provide valuable insights into how vortex beams behave in turbulent settings. The stable propagation of vortex beams through turbulence makes them promising for applications in optical communication and manipulation. As we learn more about how vortex beams interact with turbulence, their potential uses in optics and photonics continue to expand, opening up new possibilities for resilient and robust optical systems.\cite{11,12,13,14,15,16,17,18,19,20}\par

In free-space optical communication systems, the phenomenon of scintillations poses a considerable challenge, leading to heightened link error probabilities and an overall degradation in performance \cite{3}. The optimization of transmitting beams has been a focal point in extensive research endeavors aimed at mitigating the impact of scintillations on free-space optical communication systems. Various optimization techniques have been explored, encompassing modifications to the degree of coherence, degree of polarization, and the utilization of different classes of beam shapes \cite{4,5,6,7,8,9,10}. Significant strides have been made in recognizing the resilience of optical vortex beams, particularly those carrying twist phases, against the deleterious effects of turbulence when compared to optical beams lacking such helical phases \cite{11,12,13,14,15}. The inherent helical phase of vortex beams imparts orbital angular momentum (OAM), turning them into a valuable basis set for free-space optical communication channels. Leveraging these OAM modes facilitates an expansion of channel capacity without the need to increase the spectral bandwidth \cite{16}. Within the framework of a communication system, the orthogonality property of optical beams bearing vortex phases proves indispensable. This property implies that multiple independent data-carrying optical beams can be efficiently multiplexed and transmitted concurrently in free space, exponentially enhancing the system's data capacity by the total number of beams \cite{17}. While numerous studies have delved into the resilience of optical beams with vortex phases in a partially coherent regime when traversing turbulence \cite{18,19,20,21}, a notable gap persists. None of these investigations have thoroughly explored the impact of controllable dynamic turbulence within the Kolmogorov regime \cite{21,22,23,24,25,26,27,28,29,30,31,32}. This represents a crucial avenue of inquiry, as dynamic turbulence in the Kolmogorov regime introduces a layer of complexity and variability that demands comprehensive investigation \cite{33,34,35,36,37,38,39,40}. Understanding the interplay between optical beams with vortex phases and dynamically controlled turbulence in the Kolmogorov regime holds the key to advancing the robustness and efficiency of free-space optical communication systems. As technology evolves, exploring and harnessing the unique properties of optical vortex beams in the face of dynamic turbulence becomes increasingly pivotal for pushing the boundaries of communication capabilities. \cite{41,42,43,44,45,46,47,48,49,50}\par


Our study explores a new method to minimize turbulence effects on beams without using complex Adaptive Optical techniques. In the first phase, we introduced the "Post-Impacted Vortex Beam Shaping" technique, aiming to reduce turbulence impact. We simulated Kolmogorov-type turbulence in the lab using a Pseudo Random Phase Plate (PRPP). This PRPP-induced turbulence was applied to a collimated beam, creating a turbulent environment. The beam then passed through a Vortex Phase Plate (VPP), shaping it into various Laguerre-Gaussian modes. Two lenses further refined the beam's properties. Using a Charge-Coupled Device (CCD), we measured the scintillation index to assess the beam's behavior. By varying the image plane's position, we observed the beam's characteristics in different scenarios. We explored different Laguerre-Gaussian beams in this study.\par


The paper has been prepared as follows. in section 2 we have discussed theoretical and experimental procedures of our work. Section 3 contains the result analysis of our work. Finally, we have concluded in section 4.

\section{Theoretical Procedures and Mathematical Methodology}
We have presented our technique on the shaping of LG beams with different topological charges after propagating through PRPP based laboratory simulated turbulence. The turbulence follows Kolmogorov statistics. We have discussed the efficiency of turbulence noise reduction technique using pre and post beam shaping techniques.\par

\subsection{Laguerre Gaussian Beam and its Topological Charges}

Laguerre–Gaussian (LG) beams exhibit distinct physical properties that set them apart, making them highly versatile and valuable in various applications. Their unique characteristics include a barrel-shaped intensity distribution, a helical wavefront, and a center phase singularity. Furthermore, they possess the ability to carry both spin and orbital angular momentum, as well as exhibit spatial propagation invariance. These properties endow LG beams with remarkable features, making them suitable for applications such as optical pipes, optical tweezers, and optical spanners. The rotational symmetry along their propagation axis and the intrinsic rotational orbital angular momentum of $i\not{h}$ per photon further contribute to their usefulness in manipulating and controlling microscopic particles. One of the most prominent features of LG beams lies in their central vortex and partial coherence. These attributes render them highly resilient to the adverse effects of atmospheric turbulence, making them an ideal candidate for applications in challenging and turbulent environments. In particular, their ability to maintain their coherence and structural integrity amidst atmospheric turbulence is of immense importance, as it ensures stable and reliable performance in practical settings. The generation of LG beams of different orders can be achieved through various techniques. Spatial light modulators, acting as re-configurable diffractive optical elements, present one method for generating LG beams of desired orders. Additionally, Vortex phase plates offer another approach to produce LG beams with specific characteristics. Both of these methods enable precise control over the properties of LG beams, allowing for tailored applications and enhanced versatility in optical systems. Overall, LG beams represent a powerful tool in the manipulation and control of microscopic particles, owing to their distinct physical properties and resistance to atmospheric turbulence. Their ability to carry both spin and orbital angular momentum, coupled with their spatial propagation invariance, makes them a valuable asset in a wide range of optical applications. As research in this area continues to progress, the potential for LG beams to revolutionize various fields, including optical communications, micro-optical systems, and biomedical applications, becomes increasingly evident. The versatile and robust nature of LG beams opens up new possibilities for exploring novel phenomena and devising innovative solutions in the realm of optics and photonics. Usually, the Gaussian beams can be represented as follows. 

\begin{eqnarray}\label{1}
    && U (x,y,z,t)=\phi_{01}\left (\frac{W_0}{W(z)}\right ) \exp{\left (-\frac{(x^2+y^2)}{W^2 (z)}\right )}\nonumber\\&&\exp{\left (ikz-ik\frac{(x^2+y^2)}{2R(z)}+i\xi(z)-\omega t\right )}
\end{eqnarray}
The Laguerre–Gaussian modes are uniquely defined by two essential parameters, denoted as n and l, which correspond to the radial and azimuthal profiles of the beam, respectively. The parameter l holds significant importance, as it represents the orbital angular momentum carried by the beam. Furthermore, the Laguerre–Gaussian (LG) beams encompass a comprehensive set of propagation modes. These modes are characterized by the radial electric field, which can be expressed as the product of a Gaussian function and an associated Laguerre polynomial, specifically denoted as $L^{l}_{m}$. The complex amplitude of LG beam can be written as follows.
\begin{eqnarray}\label{2}
    &&U_{l,m}(\rho,\phi,z,t)=A_{l,m}\left(\frac{W_0}{W(z)} \right)\left (\frac{\rho}{W(z)} \right )^{l}L_{m}^{l}\left (\frac{2\rho^2}{W^2 (z)} \right )\nonumber\\&&\exp{\left (-jkz-jk\frac{\rho^2}{2R(z)}-jl\phi+j(l+2m+1)\xi(z)-\omega t\right )}
\end{eqnarray}

Here, $(\rho,\phi,z)$ are in cylindrical coordinates with $\rho=\sqrt{x^2+y^2}$, $k$ is the wave number, $W(z)$ is the beam width, $R(z)$ is the wavefront radius of curvature, $\xi(z)$ is the Gouy phase and $L_{m}^{l}$ is the generalized Laguerre polynomial function. The beam width is given as follows.
\begin{equation}
    W(z)=W_0 \sqrt{\left (1+\left (\frac{z}{z_0}\right )^2\right )}
\end{equation}
Where, $z_0$ is known as the Rayleigh range and $W_0$ is the waist radius. The term $R(z)$, $\xi(z)$ and $W_0$ can be written as $R(z)=z\left(1+\left (\frac{z}{z_0}\right )^2 \right )$, $\xi(z)=\tan^{-1}\left (\frac{z}{z_0} \right )$ and $W_0=\sqrt{\frac{\lambda z_0}{\pi}}$.\par
The Laguerre–Gaussian beams exhibit fascinating properties governed by two parameters, $l$ and $m$, which represent the azimuthal and radial characteristics, respectively. When $l$ and $m$ are both set to zero, the Laguerre–Gaussian beam reduces to a familiar Gaussian beam. However, as the value of $l$ increases, the electric field acquires an azimuthal phase change of $2\pi l$, leading to a phase singularity and an intensity node at the center of the beam. The parameter $l$ corresponds to the number of $2\pi$ windings around the azimuthal angle $\phi$, adding an intriguing twist to the beam's characteristics [7]. The Laguerre–Gaussian beam displays circular symmetry in its intensity profile, dependent on the variables $\rho$ and $z$. For $l\neq 0$, the beam exhibits zero intensity at its center $(\rho=0)$ and takes on an annular intensity pattern. The phase $\phi(\rho,z)$ shares similarities with the Gaussian beam, but there are noteworthy differences. Specifically, the Gouy phase is greater by a factor of $(l+2m+1)$, and an additional term proportional to the azimuthal angle $\phi$ is present. Due to this linear dependency on the azimuthal angle $\phi$ (for $l\neq 0$), the wavefront of the beam assumes a helical form as it propagates in the $z$ direction. Such beams with a spiral phase are of particular interest due to their capability to carry orbital angular momentum, enabling them to impart torque to illuminated systems [7]. The Laguerre–Gaussian beams possess an essential property of axial symmetry and feature spherical wavefronts. These beams naturally serve as eigen modes of optical systems characterized by spherical optical surfaces and cylindrical symmetry. Their unique properties and ability to carry orbital angular momentum make them intriguing and versatile tools for various applications in optics and related fields [7].

\subsection{Conversion from Symmetric Gaussian Wavefront to Vortex Gaussian Wavefront}
In our present work, we discuss the vortex beam with radial index $m=0$. Hence, we don't need to consider the radial phase factor generated due to presence of non-zero $m$. We have introduced vortex phase plate (VPP) to convert the symmetric Gaussian beam into rotational symmetric vortex beams by addition of azimuthal phase factor on the input beam. The VPP phase can be shown schematically as follows.

\begin{figure}[H]
\centering
\begin{minipage}[b]{0.45\textwidth}
    \includegraphics[width=\textwidth]{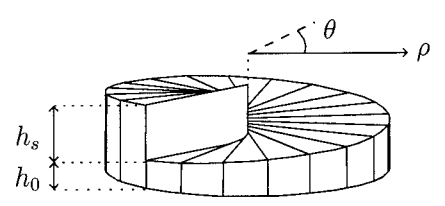}
    \caption{Pictorial representations Vortex Phase Plate}
    \label{fig:2}
\end{minipage}
\hfill
\begin{minipage}[b]{0.45\textwidth}
    \includegraphics[width=\textwidth]{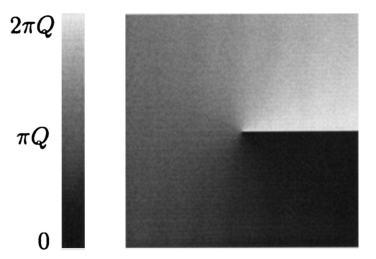}
    \caption{Phase distribution imprinted onto the transverse plane of an incident beam}
\label{fig:3}
\end{minipage}
\end{figure}

A spiral phase plate is an optical device characterized by a transparent plate with a thickness that increases proportionally to the azimuthal angle $\phi$ around a central point. This arrangement resembles a single turn of a staircase, where the step height is denoted as $h_{final}=(h_s +h_0)$. The fundamental principle behind its operation lies in imparting a helical character to a light beam that traverses through it. In a rigorous analysis, the intricate behavior of the plate necessitates the application of vector-diffraction theory. However, within the paraxial regime and for beams with limited divergence, as well as a sufficiently small step height, a simplified treatment considering the operation solely on the phase becomes feasible. When a beam interacts with the spiral phase plate, it undergoes transformations that are especially pronounced for beams with a small divergence. The plate can convert a non-helical beam into a helical one, showcasing its capacity to induce helicity changes. It is important to note that this transformation may not only generate helical beams but also alter the helicity of an initially helical beam. The modification of helicity, as expressed by equation 4, corresponds to a change in the phase distribution while leaving the amplitude distribution unchanged. However, this modification doesn't always yield a pure mode that propagates with a simple rescaling of the intensity distribution. Instead, the resulting beam tends to be a superposition of modes. A Spiral Phase Plate (SPP) functions as an optical component introducing azimuth-dependent retardation to the optical field. Its basic configuration, illustrated in Figure 2, involves a transparent plate with a refractive index $n$, where the plate's height is directly proportional to the azimuthal angle $\theta$.
\begin{equation}
    h=h_s \frac{\theta}{2\pi}+h_0
\end{equation}
where $h_s$ is the step height and $h_0$ is the base height of the device. This configuration yields an azimuth dependent optical phase delay;
\begin{equation}
    \phi(\theta,\lambda)=\frac{2\pi}{\lambda}\left [\frac{(n-n_0)h_s \theta}{2\pi}+nh_0\right ]
\end{equation}
In the presence of a spiral phase plate, characterized by its refractive index $n$ and inserted within the waist of a Gaussian beam where the phase distribution is planar, a vortex of charge $Q=\frac{h_s(n-n_0)}{\lambda}$ is imparted. Here, $n_0$ represents the refractive index of the surrounding medium, $\lambda$ is the wavelength of the incident light, and $h_s$ is the step height of the spiral phase plate. This process results in an output beam with an associated orbital angular momentum per photon of $Q\not{h}$. Importantly, the value of $Q$ relies explicitly and implicitly, through the refractive index, on the wavelength of the incident light. This characteristic underscores the chromatic nature of the spiral phase plate device. The resultant phase distribution, as a consequence of these transformations, is visually depicted in Figure 3. The application of such phase plate can change the direction of output electric fields to diverge it when collimated fields propagate through it. If we consider a plane wavefront with Gaussian spatial intensity distribution which is propagated through a vortex phase plate, using equation \ref{1} should be written as follows with $k\frac{(x^2+y^2)}{2R(z)}=\xi(z)=0$, $W_0=W(z)$ and radial index $m=0$ with azimuthal index $l\neq 0$.
\begin{equation}\label{6}
    U (x,y,z,t)=\phi_{01}\exp{\left (-\frac{(x^2+y^2)}{W^2_0}\right )}\exp{\left (ikz-\omega t\right )}
\end{equation}
Hence, after passing through VPP we can find the following vortex field as output before using any lens combinations for providing shearing on the wavefront where we have $\rho=\sqrt{x^2+y^2}$.
\begin{equation}\label{7}
    U_{l,0} (\rho,\phi,z,t)=\phi_{01}\exp{\left (-\frac{\rho^2}{W^2_0}\right )}\exp{\left (ikz-\omega t\right )}\exp{(il\phi)}
\end{equation}
After application of a lens with focal length $f_1$ we can find the Fourier transform of the field eqn.\ref{7} on the focal plane of the lens. If we introduce another lens with focal length $f_2$ at a distance of $f_1+f_2$ from first lens, we can find inverse Fourier transform distribution of the wavefront of first focal plane. The intensity distribution on first focal plane can be written as follows with azimuthal phase factor $\theta$, radial coordinate $\rho'$ and $a=\frac{k^2W_0^2}{8\omega^2}$ where, $\lambda=$wavelength and $k=$ wave vector.
\begin{eqnarray}
    &&U'_{l,0} (\rho',\theta,z,t)=i^l \sqrt{\pi}W_0\omega a \rho'\exp{(-a\rho'^2)}\nonumber\\&&\left [L_{\frac{l-1}{2}}(a\rho'^2)-L_{\frac{l+1}{2}}(a\rho'^2) \right]\exp{\left (ikz-\omega t\right )}\exp{(il\theta)}
\end{eqnarray}
Here, $L$ represents the modified Bessel function of first kind. The wavefront distribution carries the information of spatial vorticity on Fourier plane. Now, the inverse Fourier transformation can be written as follows.
\begin{eqnarray}
    &&U''_{l,0} (\rho,\phi,z,t)=(-1)^l (2a)^{2/3}\pi^{1/2}\exp{\left (ikz-\omega t\right )}\nonumber\\&&\exp{(il\phi)}\int_{0}^{\infty}\rho'^2\exp{(-a\rho'^2)}\left [L_{\frac{l-1}{2}}(a\rho'^2)-L_{\frac{l+1}{2}}(a\rho'^2) \right]\nonumber\\&&J_l \left (\frac{2\pi\rho'\rho}{\omega\lambda}\right )d\rho'
\end{eqnarray}

\subsection{Pseudo Random Phase Plate: Mathematics of Beam Patches}
The pseudo random phase plates works as a random refracting optical components which can produce different electric displacement vectors when any continuous wavefront propagated through it. Due to variation of total optical paths, each pixel of the wavefront gets different directions of their displacement vectors. This phenomena causes discontinuous patches on the initial continuous wavefront. If we establish a dynamic configuration of the PRPP, the discontinuous patches movements can be observed. This condition is called as dynamic turbulence which follows Kolmogorov statistics. Using this technique we can say that the wavefronts which remains in complex plane can be effected more that the wavefronts on real plane. When a beam with complex degrees of freedom propagated through PRPP, ellipticity of the polarization state gets changed because of getting different phases on different basis states of the initial polarization states. This phenomena can be incarnated on LG beams. In general LG beams are the amplitude distribution on complex plane. Hence, the LG beams get higher turbulence impact when they are propagated through PRPP. On the other hand, if we shape a Gaussian beam which is propagated through turbulence, the turbulence impact will be less. These phenomena have been established in experiments as follows. The PRPP provides inhomogeneous optical path difference which can be represented as follows. Here, the two refractive indices are $n_1$ and $n_2$ with variable path $h_1(x,y,z)$ and $h_2(x,y,z)$. \textbf{(Figure 1)}
\begin{equation}
    \Delta (x)=h_2(x,y,z)n_2 + h_1(x,y,z)n_1
\end{equation}
The trajectory of electromagnetic field through any dielectric medium can be discussed with the electric displacement vectors of the light beam field. Mathematically, if we consider the input electric field of the beam as $E_i$ and vacuum permittivity $\epsilon_0$, the electric displacement vector can be written as follows.
\begin{equation}
    \overrightarrow{D} = \epsilon_0 \overrightarrow{E_i} + \overrightarrow{P}
\end{equation}
Here, $\overrightarrow{P}$ is the polarization vector which comes from the dipoles created when the field starts its propagation through the medium. Here we assumed that the quantity polarization acts as a vector inside the PRPP. This can be related with the vacuum permittivity $\epsilon_0$ using electric susceptibility $\chi$ as follows.
\begin{equation}
    \overrightarrow{P}=\epsilon_0 \chi \overrightarrow{E_1}
\end{equation}
Hence, \textbf{equation 11} can be written as $\overrightarrow{D} = \epsilon_0 (1+\chi) \overrightarrow{E_i}$ which represents that the direction of propagation of electric displacement vector is totally dependent upon the direction of propagation of input electric field. We can again say that the refractive index of a dielectric material is dependent upon electric permittivity as $n=\sqrt{\epsilon}$. Thus, for the two medium inside PRPP, the electric displacement vectors through them can be written as follows. Here, the input field $E_i$ is for first medium and $E_f$ is for second medium which is also the output field of first medium. We have taken $\epsilon_0 (1+\chi_1)=\epsilon_1=n_{1}^{2}$ and $\epsilon_0 (1+\chi_2)=\epsilon_2=n_{2}^{2}$ for those two medium.
\begin{equation}
    \overrightarrow{D_1} = \epsilon_0 (1+\chi_1) \overrightarrow{E_i}
\end{equation}
and;
\begin{equation}
    \overrightarrow{D_2} = \epsilon_0 (1+\chi_2) \overrightarrow{E_f}
\end{equation}
Now, as we know the input surface of the first medium is planner in nature and hence, all the displacement vectors of the beam propagate in same direction. The thickness of the first medium is represented by $h_1 (x,y,z)$ which provides different optical path difference because of the factor $h_1 (x,y,z)\sqrt{\epsilon_0 (1+\chi_1)}$. On the other hand, the input plane of the second medium is non-uniform because of the factor $h_1(x,y,z)$ which can produce different propagation directions of the displacement vectors through second medium. This phenomena can discritize the initial continuous Gaussian intensity distribution which provides the patches on the wavefront. The thickness of the second medium i.e. $h_2 (x,y,z)\sqrt{\epsilon_0 (1+\chi_2)}$ can also give different optical path differences for different pixels. The final output plane of the second medium is planner and uniform which can't give any additional effect. Hence, the final output wavefront coming out from second medium can be observed as discrete wavefront patches. The dynamic nature of the PRPP can include time or rotation speed dependency on the optical path differences and non-uniformities of medium planes. Hence, the patches positions can move with respect to final output image plane. Thus, the rotating PRPP can produce turbulence which follows Kolmogorov statistics.

\subsection{Polarized light Propagation through PRPP Turbulence}
Whenever a plane polarized light is propagated through turbulence, the state of polarization remains unaltered. Due to having randomness in the phases of PRPP, the state may get some addition phases but the oscillation direction of the beam fields remain unchanged. This can be written as follows mathematically with the phase factor $\theta_1$ and $\theta_2$.
\begin{equation}
    \left | H_f\right > = e^{i\theta_1}\left | H_i\right >
\end{equation}
and;
\begin{equation}
    \left | V_f\right > = e^{i\theta_2}\left | V_i\right >
\end{equation}
Without loss of generality we must have to choose $\theta_1\neq\theta_2$. Hence, for ellipticity the final state can be represented as follows.
\begin{equation}
    \left |\Psi_{ellipse}\right > = \alpha_1 \left | H_f\right >+i\alpha_2e^{i\phi_f} \left | V_f\right > = \alpha_1 e^{i\theta_1}\left | H_i\right >+i\alpha_2e^{i\phi_i} e^{i\theta_2}\left | V_i\right >
\end{equation}
If we consider $\alpha_1 \left | H_i\right >+i\alpha_2e^{i\phi_i} \left | V_i\right >=\left |\psi_{ellipse}\right >$, we can observe that the ellipticity has been changed from $\left |\psi_{ellipse}\right >$ to $\left |\Psi_{ellipse}\right >$ conversion through turbulence medium. Hence, we can say that the elliptic polarized states are effected with PRPP based turbulence medium in terms of their ellipticity measurements. In another way, we can say that the wavefronts in complex plane can be effected with turbulence due to having different phases for different basis states through PRPP medium.

\subsection{Complex Wavefront Propagation through PRPP Turbulence}
Whenever we'll introduce LG beams, the wavefronts should work in complex plane. We can represent the LG beam modes (ex. LG1,LG2,LG3 and so on in angular momentum basis) in terms of the HG modes (Hermite Gaussian modes) in complex planes. The concept of the orbital angular momentum (OAM) Poincaré sphere is elaborated upon through the analysis of transformations between Laguerre Gaussian (LG) and Hermite-Gaussian (HG) modes in a laser beam. Within this framework, a LG mode characterized by zero radial index and an azimuthal index denoted as "l" signifies an optical vortex of a specific order "l". This mode can be equivalently described in the context of HG modes, revealing the intricate relationship between LG and HG representations. Here, $\begin{bmatrix}
        l \\
        s
    \end{bmatrix}$ is the binomial coefficient.
\begin{equation}
    LG^{l}_{0}=\frac{1}{2^l}\sum^{l}_{s=0}\begin{bmatrix}
        l \\
        s
    \end{bmatrix}i^s HG_{l-s,s}
\end{equation}
Now, if we choose $\left |\psi\right >$ as initial LG beam and $\left |\Psi\right >$ to represent final LG beam after propagation through turbulence, we can say that the different HG modes will get different phases due to presence of turbulence. Hence, after turbulence the final wavefront should be different from usual LG modes. This phenomena can't happen for a wavefront on real plane viz HG beams. each HG beams can get different unique phases which can't provide any change of the wavefront distribution. Hence, we can say that the wavefront in complex plane can be effected more than the wavefront in real plane when they are propagating through PRPP. Whenever, the beams are shaped after turbulence impact, none of the efore mentioned phenomena can happen and that's why, the scintillation index for turbulence impacted beam can be reduced with LG conversion after turbulence impact.\par

\subsection{Kolmogorov Statistics for Turbulence}
Classical turbulence, a phenomenon rooted in the unpredictable fluctuations of velocity in a viscous fluid like the atmosphere, manifests itself in the atmospheric motion, featuring two discernible states: laminar and turbulent. Laminar flow lacks the dynamic mixing found in turbulent flow, where random subflows, termed turbulent eddies, play a crucial role. The Reynolds number $(Re)$, defined as $Vl/\nu$, with $V$ as velocity (speed), $l$ as the flow dimension, and $\nu$ as kinematic viscosity, determines the transition from laminar to turbulent flow. The critical Reynolds number for turbulent conditions, especially near the ground, is typically around $10^5$, signifying highly turbulent behavior. Kolmogorov's turbulence theory advances a set of hypotheses asserting that small-scale structures exhibit statistical homogeneity, isotropy, and independence from large-scale structures. The energy source at large scales is attributed to either wind shear or convection. As wind velocity surpasses the critical Reynolds number, large unstable air masses emerge, initiating the process of energy cascade theory. This theory posits that under the influence of inertial forces, unstable air masses disintegrate into smaller eddies, establishing a continuum of eddy sizes for the transfer of energy from a macro-scale $L_0$ (outer scale of turbulence) to a micro-scale $l_0$ (inner scale of turbulence). Within this framework, an inertial range emerges, defined by eddies bounded by $L_0$ above and $l_0$ below. This range represents a critical domain where energy transfer through eddies occurs. Beyond the inertial range lies the dissipation range, encompassing scale sizes smaller than $l_0$. In this domain, the remaining energy in fluid motion dissipates as heat, marking the conclusion of the energy cascade process. Understanding classical turbulence involves recognizing these intricate scales and the interplay between large and small structures. The dynamic transition from laminar to turbulent states, guided by critical Reynolds numbers, highlights the inherent complexity of turbulent flows in the atmosphere, a phenomenon crucial for comprehending various natural processes and phenomena.The power spectrum for this type of turbulence can be written as follows.
\begin{equation}
    \Phi(k)=0.023r_0^{-5/3}k^{-11/3}
\end{equation}

\subsection{Scintillation Index and Relative Noise Mapping}
One of the primary parameters used to quantify the impact of turbulence on optical beams is known as the scintillation index. This index serves as a measure of the intensity fluctuations that arise due to the refractive index fluctuations occurring in the turbulent atmosphere. Scintillations refer to the variations in intensity, and they significantly affect the quality of optical beams. When a propagating beam encounters atmospheric turbulence, these scintillations can cause undesirable effects such as beam wandering, beam spreading, and intensity variations. As a consequence, the overall performance of optical systems, such as ground-based telescopes and free-space optical communication systems, can be severely compromised by the presence of scintillations. Understanding and characterizing the scintillation index are crucial for mitigating the adverse effects of atmospheric turbulence on optical beam propagation and for devising effective methods, such as adaptive optics, to compensate for the scintillation-induced distortions. The scintillation index has become a key parameter in evaluating and optimizing the performance of optical communication links and other optical systems operating in turbulent environments. Moreover, research efforts continue to explore novel techniques and technologies that can further minimize the impact of scintillations, thus enhancing the reliability and robustness of optical beam transmission and manipulation in challenging atmospheric conditions.\par
For laser beam the scintillation index is measured with the relative difference between the square of RMS and Mean of the intensities of images with respect to the square of Mean. The functional form can be written as follows.
\begin{equation}
    \sigma^2=\frac{\langle I^2\rangle-\left (\langle I\rangle\right )^2}{\left (\langle I\rangle\right )^2}
\end{equation}
Where, $I$, $\langle I\rangle$ and $\langle I^2\rangle$ are respectively the instantaneous intensity, the average intensity, and the intensity correlation function of the beam in the detector plane. \textbf{In experimental result analysis, the scintillation has been defined pixel wise over all the frames and plottings have been done with average scintillation index over all pixels. Hence, equation 20 can be used as the experimental form as $\left <\sigma^2\right >_{ij}=\left <\frac{\langle I^2\rangle_{N}-\left (\langle I\rangle_N\right )^2}{\left (\langle I\rangle_N\right )^2}\right >_{ij}$. Here, $\left <\right >_{ij}$ represents the average over all pixels and $\left <\right >_N$ represents average over all frames.} Another way to study the phase fluctuation and noises can be represented with beam wandering. The scheme can be written as follows.
\begin{equation}
    \sqrt{\left (\left <r_c^{2}\right >\right )}=\sqrt{\frac{(x_c-<x_{c}>)^2+(y_c-<y_{c}>)^2}{N}}
\end{equation}
Here, $N$ is the total number of frames. $x_c$ and $y_c$ are the coordinates of the centroids of the frames. The centroid can be defined as follows.
\begin{equation}
    x_c =\frac{\sum_{j=1}^{J}\sum_{i=1}^{I}x_i I_{ij}}{\sum_{j=1}^{J}\sum_{i=1}^{I} I_{ij}}
\end{equation}
And,
\begin{equation}
    y_c =\frac{\sum_{j=1}^{J}\sum_{i=1}^{I}y_i I_{ij}}{\sum_{j=1}^{J}\sum_{i=1}^{I} I_{ij}}
\end{equation}
Where, $I$ and $J$ are the total number of pixels.

\section{Experimental Procedures and Results}
The Pseudo Random Phase Plate (PRPP) utilized in our experimental setup to simulate atmospheric turbulence was provided by Lexitex Corporation, USA \cite{28,29,30,31}. This PRPP is a multi-layered device consisting of five layers, where two outer optical windows made of BK7 glass are sandwiched between two inner layers of near-index-matching polymer, and a single layer of acrylic containing the desired turbulence profile is imprinted on one side \cite{28}. To achieve near-index matching (NIM), a single layer of polymer is placed on each side of the acrylic, resulting in a mechanically stable configuration. The PRPP is further sealed with optical window glass to ensure the structural integrity and minimize the impact of environmental factors, such as thermal expansion-induced stress, on the phase plate's performance. The PRPP has a thickness of approximately 10 mm, making it mechanically robust and easily mountable on a rotary stage for precise control. The entire Pseudo Random Phase Plate (PRPP) was manufactured by Lexitek Corporation, USA \cite{32,33,34,35,36,37,38,39,40}, and the design was developed by Starfire Optical Range, USA. The PRPP used in our experiment is specifically imprinted with Kolmogorov type turbulence at optical wavelengths. Detailed illustrations of the PRPP used in the experiment can be found in Fig. 1 a, b. By adjusting the turbulence strength, the PRPP allows us to simulate aberrated wavefronts with known Fried coherence lengths, denoted as $r_0$, ranging from 16 to 32 samples. This enables us to select various levels of turbulence to accurately mimic different atmospheric conditions. Each phase screen contains 4096 sample phase points across one side. The turbulence profile is mapped onto a 3.28’’ acrylic annulus with a 1.35’’ diameter obscuration at the center, ensuring a spacing of 20 $\mu$m at each plate \cite{28}(figure \ref{Fig: 1}). \par

\subsection{Description of Experimental Setup}
The experimental geometry used in our experiment is shown in Fig.4 in which a He-Ne laser of wavelength 6328 A is used as the source. The unexpanded laser beam from the laser is spatially filtered and passes through the rotating Pseudo Random Phase Plate (PRPP) which creates the dynamic turbulence. The Vortex Phase Plate(VPP) shown in Fig.4 at positions respectively before and after PRPP is used for converting laser beams  to topologically charged Laguerre Gaussian beams before and after passing through turbulence. The experimental analysis is carried out at planes located in 5 different planes(Fig.4) by keeping the CCD camera at all these planes. The turbulence impacted beams coming from the PRPP passes through a 4f imaging system as shown in Fig.4 in which the first location of measurement plane 1 is just before the first lens(Lens 3) of the 4f imaging system. The second location of measurement plane(Plane 2) is kept immediately after the lens 3 (converging position), the third plane(Plane 3) is the Fourier Transform plane of Lens 3 (First lens of 4f imaging system), then the plane 4 is located immediately after the focal plane of lens 3 along the diverging beam and finally the 5th plane(Plane 5) is located at the inverse Fourier Transform plane in the 4f imaging system as shown in Fig.4. At all these 5 positions of planes 30 images with pixel dimensions 1024 X1024 are taken for beams with and without turbulence impact. \par

\begin{figure}[H]
\centering
\begin{minipage}[b]{1.0\textwidth}
    \includegraphics[width=\textwidth]{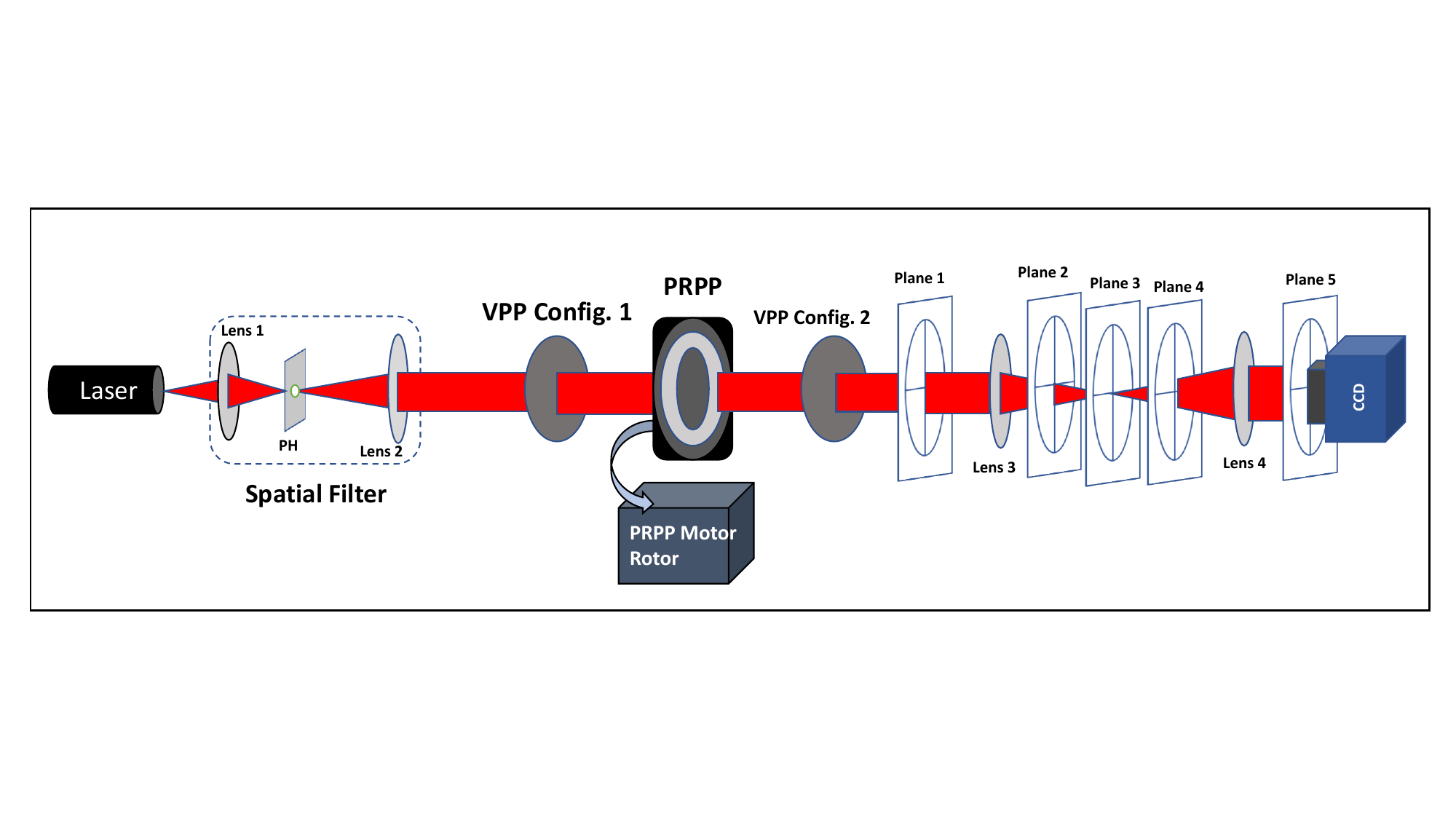}
    \caption{Experimental diagram of turbulence impacted beam shaping}
    \label{2}
\end{minipage}
\end{figure}

\subsection{Results and Discussion}
At all these 5 planes, the scintillation index and beam wandering along radial ($\sqrt{\left <r_c^2\right >}$) direction is measured and plotted for simple Gaussian beams, Laguerre Gaussian(LG) beams of 4 topological charges with and without turbulence impact. Also, the beam shaping is carried out by converting LG beams post turbulence impact and their respective scintillation index and beam wandering characteristics along radial ($\sqrt{\left <r_c^2\right >}$) direction at all these 5 planes are carried out and plotted. The results recorded at plane 1 (before 4f lens imaging system as shown in Fig.4) is shown in Fig.5, where it shows the graph plotted with beam modes along x axis and Mean Scintillation index along y axis respectively. The blue line shows the plot when the Laguerre Gaussian beams(LG) was propagated without turbulence, The red line shows the plot for LG beams shaped(converted) after the impact of turbulence(VPP at a position after the PRPP, Fig.4)  and the yellows showing the plot for LG beams propagated through the turbulence(VPP at a position before PRPP, Fig.4). The plots shown in Fig.6 is measured at the plane 2 that is immediately after the lens 3 which is along converging position in the 4f imaging system(Fig.4). Similaraly, Fig .7, Fig.8, Fig.9 show the plots obtained at planes 3, 4 and 5 respectively. It can be seen from these plots the best result is obtained for all the turbulence impacted LG beams post conversion(shaping) i.e when VPP is kept after the PRPP(Fig.4) for beam shaping(red line) compared to LG beams with turbulence impact(green line). This result clearly shows that when the turbulence impacted beam is converted or shaped to different LG beams with different charges from 1 to 4) we can obtain less scintillation index and the values are uniformly same as the redline is almost a straight line. Also, one can easily see from these plots the LG beams shaped after the turbulence impact show less scintillation index values at all planes except when the measurement is taken at plane 1 i.e before the 4f lens imaging system. \par

\begin{figure}[H]
\centering
\begin{minipage}[b]{0.45\textwidth}
    \includegraphics[width=\textwidth]{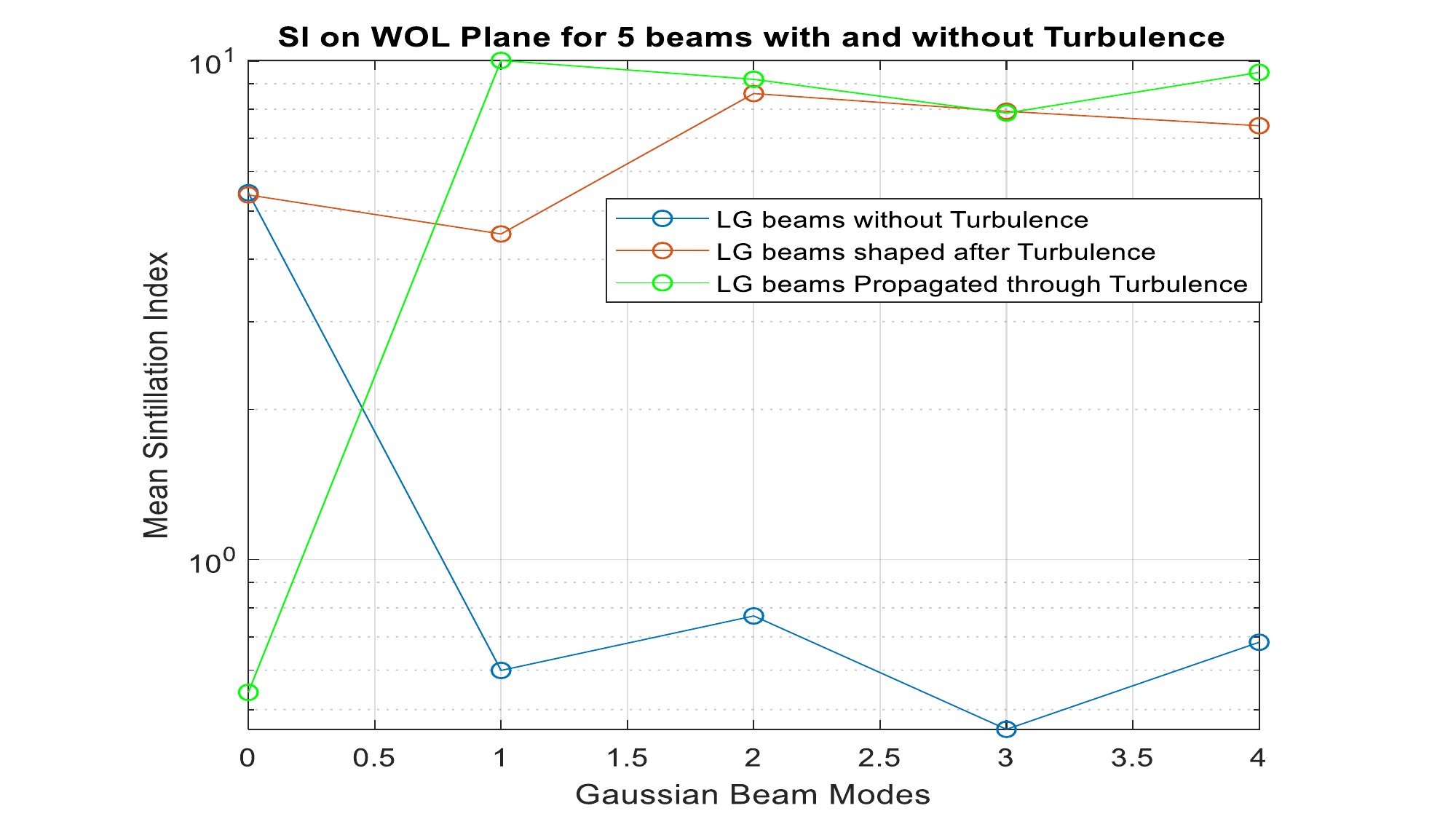}
    \caption{\textbf{SI Comparison for different Beams; Blue: LG beams propagated without Turbulence; Red: LG beams shaped after Turbulence; Green: LG beams propagated through Turbulence for Plane 1}}
    \label{Result:1}
\end{minipage}
\hfill
\begin{minipage}[b]{0.45\textwidth}
    \includegraphics[width=\textwidth]{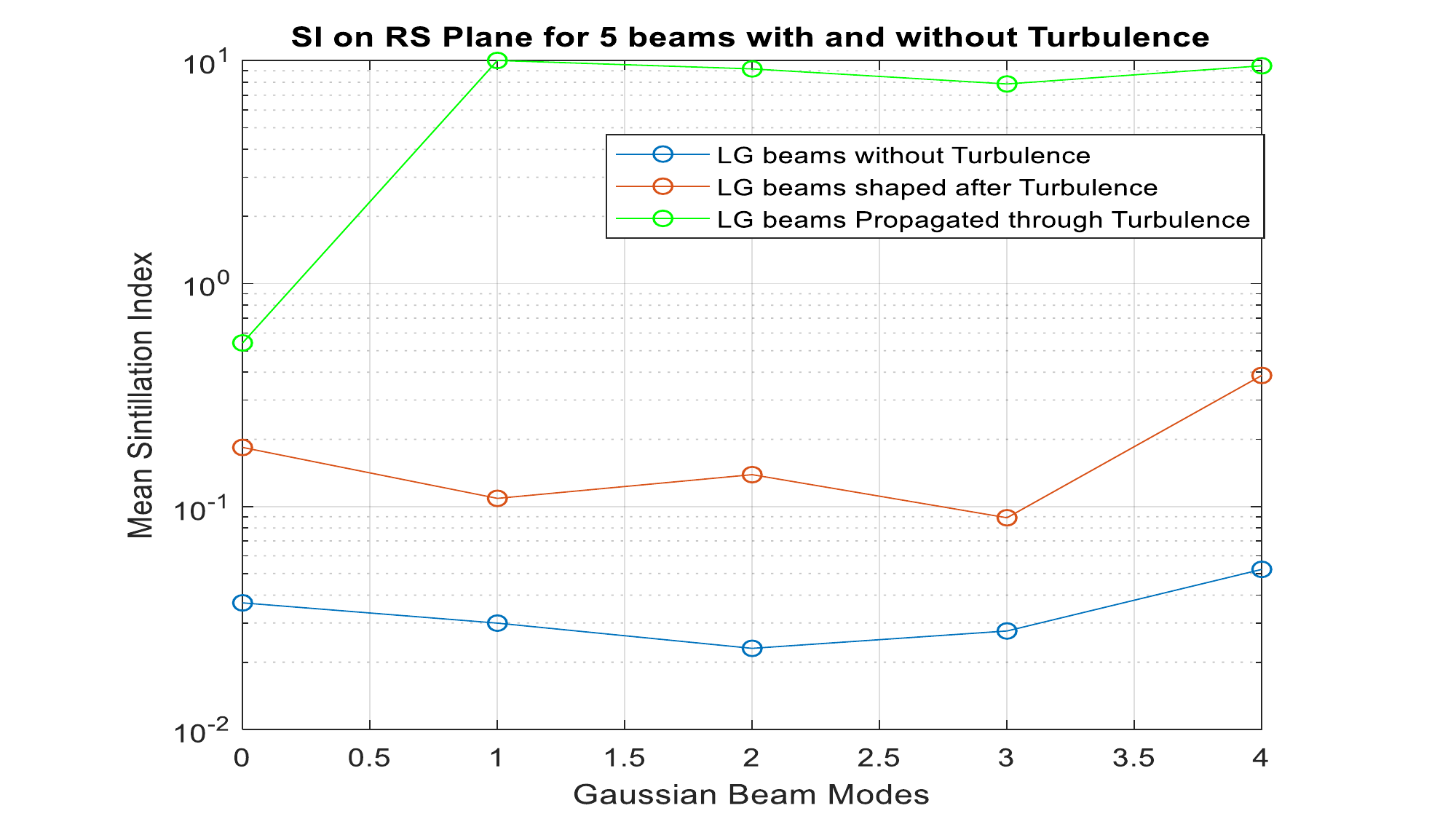}
    \caption{\textbf{SI Comparison for different Beams; Blue: LG beams propagated without Turbulence; Red: LG beams shaped after Turbulence; Green: LG beams propagated through Turbulence for Plane 2}}
    \label{Result:2}
\end{minipage}
\end{figure}

\begin{figure}[H]
\centering
\begin{minipage}[b]{0.45\textwidth}
    \includegraphics[width=\textwidth]{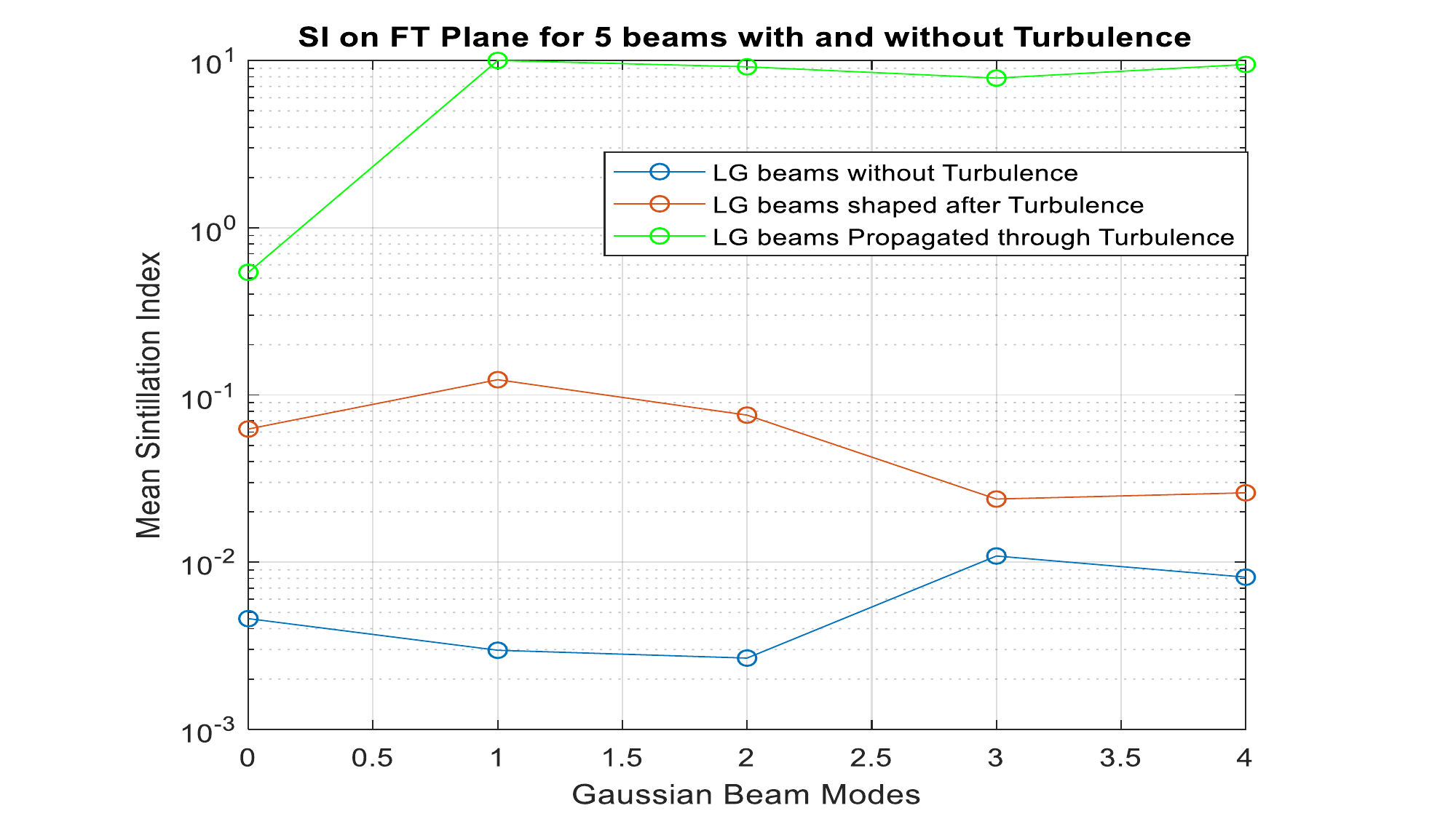}
    \caption{\textbf{SI Comparison for different Beams; Blue: LG beams propagated without Turbulence; Red: LG beams shaped after Turbulence; Green: LG beams propagated through Turbulence for Plane 3}}
    \label{Result:3}
\end{minipage}
\hfill
\begin{minipage}[b]{0.45\textwidth}
    \includegraphics[width=\textwidth]{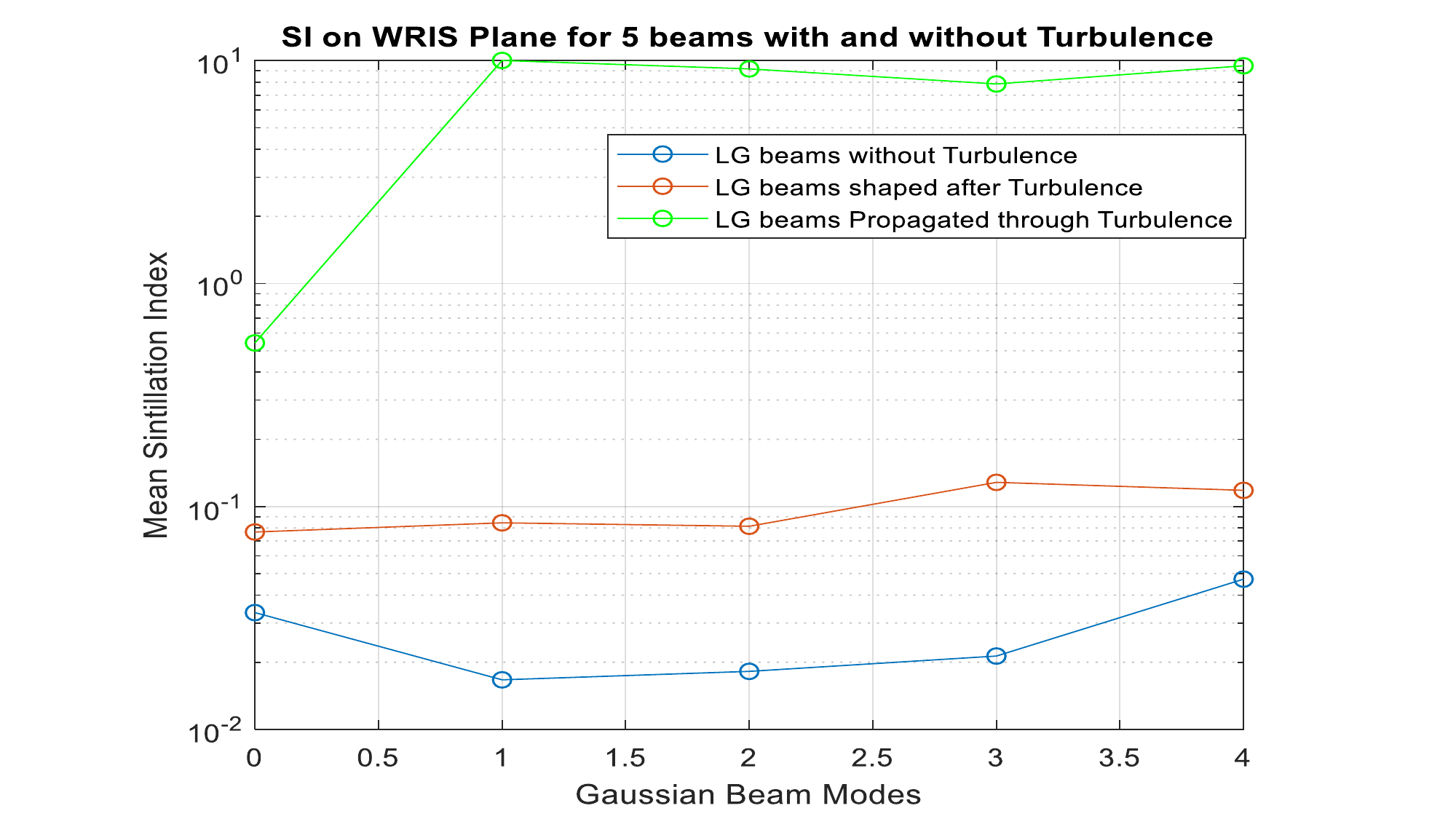}
    \caption{\textbf{SI Comparison for different Beams; Blue: LG beams propagated without Turbulence; Red: LG beams shaped after Turbulence; Green: LG beams propagated through Turbulence for Plane 4}}
    \label{Result:4}
\end{minipage}
\end{figure}

\begin{figure}[H]
\centering
\begin{minipage}[b]{0.45\textwidth}
    \includegraphics[width=\textwidth]{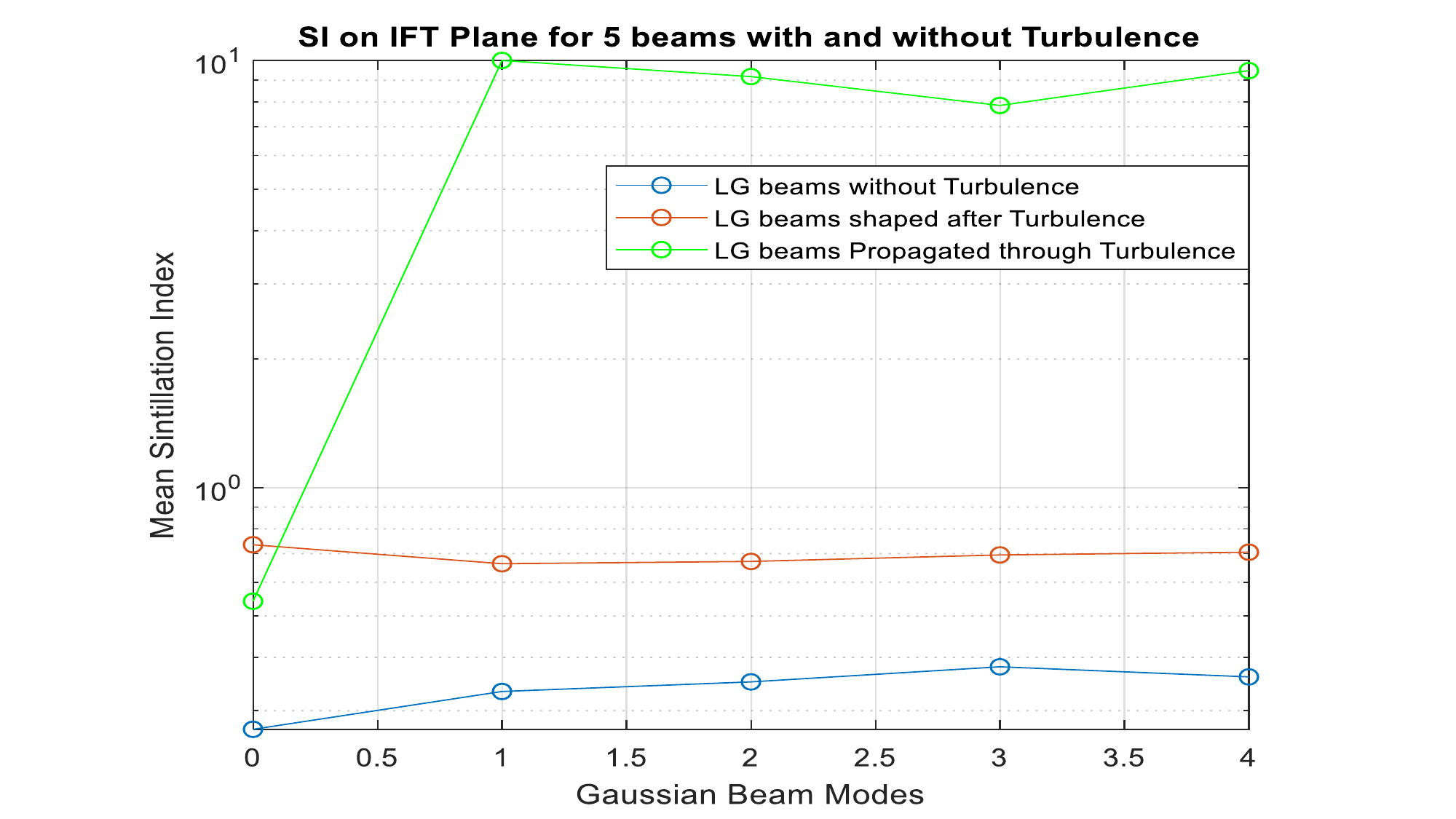}
    \caption{\textbf{SI Comparison for different Beams; Blue: LG beams propagated without Turbulence; Red: LG beams shaped after Turbulence; Green: LG beams propagated through Turbulence for Plane 5}}
    \label{Result:5}
\end{minipage}
\end{figure}
We have further plotted graphs by keeping positions of different planes along x-axis and mean scintillation index values along y axis for the LG beams having different topological charges propagating through turbulence and the turbulence impacted beams converted on to different topologically charged LG beams (shaped after the turbulence impact).  Fig. 10, 11,12 and Fig.13 show the graphs plotted for LG beams with charge 1, 2,3 and 4 respectively for turbulence impact before and after passing through PRPP and measured at 5 different planes. In all these graphs, the red lines show the values obtained for LG beams of topological charges from 1 to 4, propagated through turbulence and blue line graphs show the plot obtained at 5 different planes when the turbulence impacted beams are shaped to LG beams of different topological charges respectively. From these plots it is easily seen that the mean scintillation index values are much less for the laser beams shaped or converted on to LG beams after the turbulence impact compared to the LG beams propagated through the turbulence. This is new observation compared to earlier results reported in the literature. \par

\begin{figure}[H]
\centering
\begin{minipage}[b]{0.45\textwidth}
    \includegraphics[width=\textwidth]{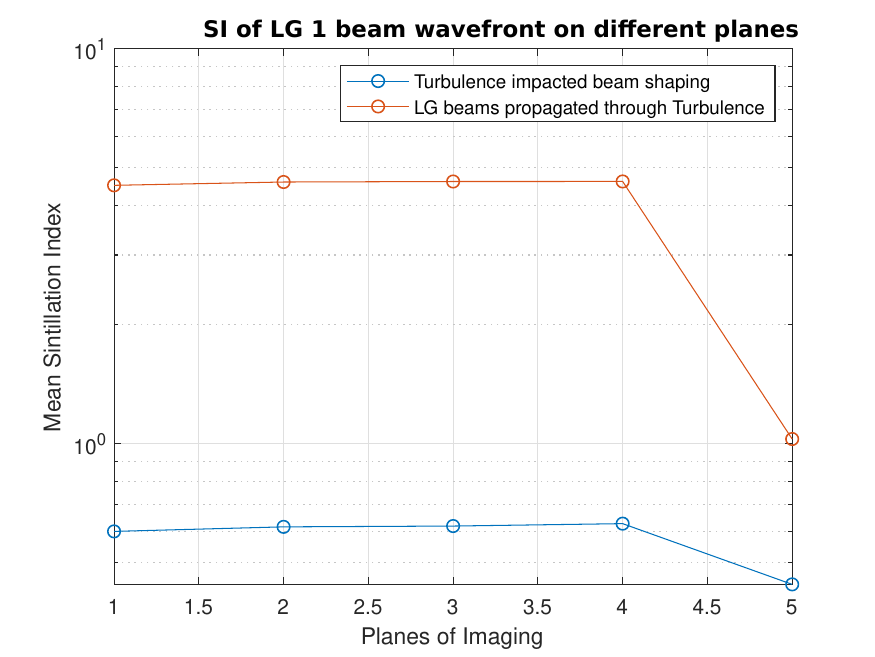}
    \caption{Scintillation Index (SI) for Laguerre Gaussian beam azimuthal charge 1 wavefront on different 5 planes. Blue curve is measurement for turbulence impacted beam shaping. Red curve is measurement with LG beams propagated through turbulence. The integer values on horizontal axis represents the position of planes according to experimental set up diagram}
    \label{Result:22}
\end{minipage}
\hfill
\begin{minipage}[b]{0.45\textwidth}
    \includegraphics[width=\textwidth]{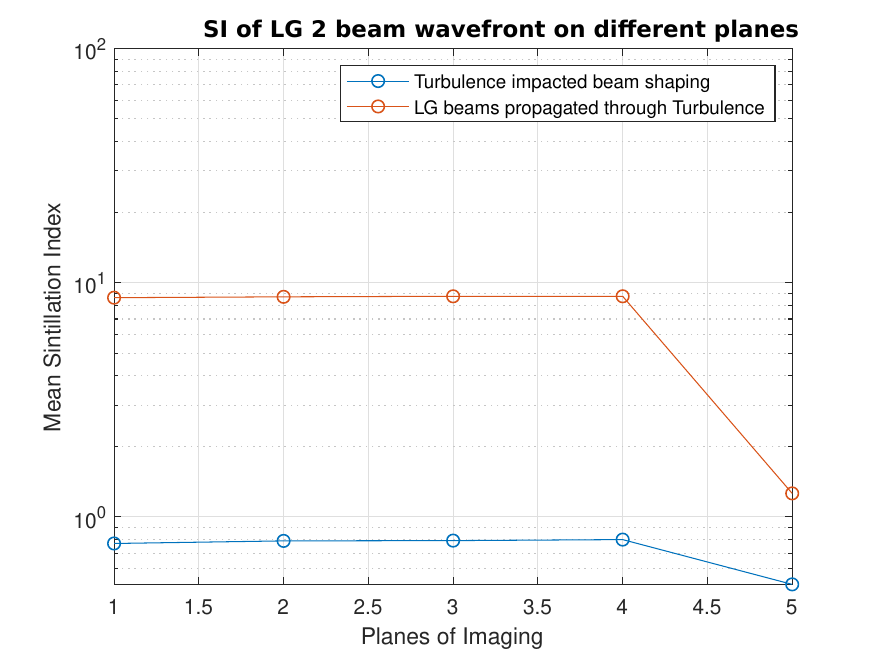}
    \caption{Scintillation Index (SI) for Laguerre Gaussian beam azimuthal charge 2 wavefront on different 5 planes. Blue curve is measurement for turbulence impacted beam shaping. Red curve is measurement with LG beams propagated through turbulence. The integer values on horizontal axis represents the position of planes according to experimental set up diagram}
    \label{Result:33}
\end{minipage}
\end{figure}
\begin{figure}[H]
\centering
\begin{minipage}[b]{0.45\textwidth}
    \includegraphics[width=\textwidth]{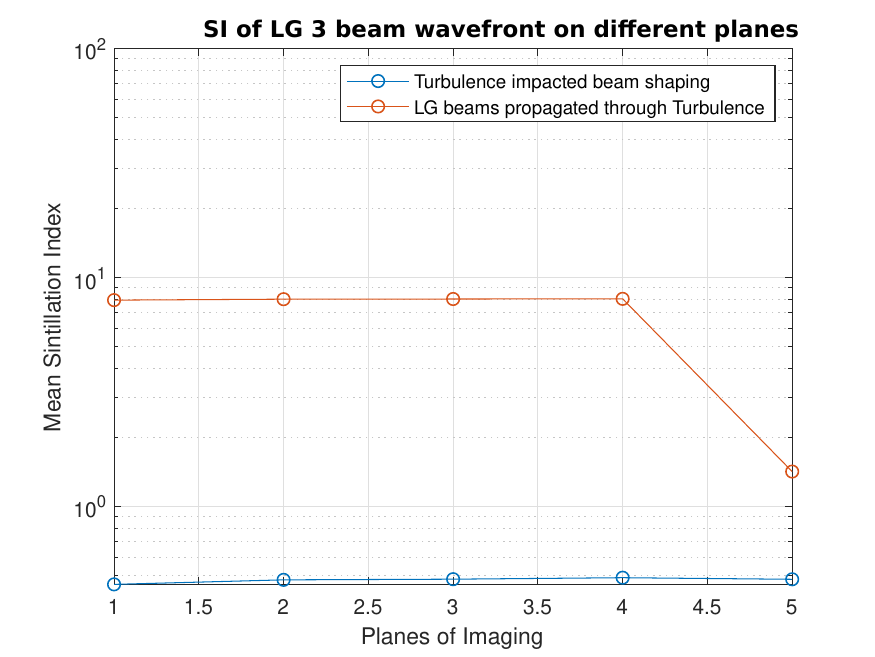}
    \caption{Scintillation Index (SI) for Laguerre Gaussian beam azimuthal charge 3 wavefront on different 5 planes. Blue curve is measurement for turbulence impacted beam shaping. Red curve is measurement with LG beams propagated through turbulence. The integer values on horizontal axis represents the position of planes according to experimental set up diagram}
    \label{Result:44}
\end{minipage}
\hfill
\begin{minipage}[b]{0.45\textwidth}
    \includegraphics[width=\textwidth]{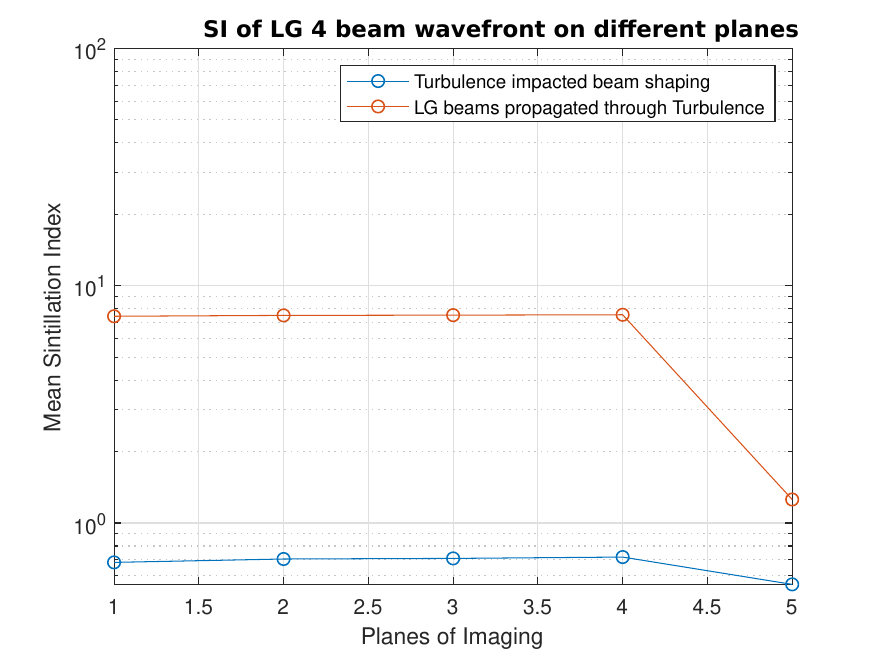}
    \caption{Scintillation Index (SI) for Laguerre Gaussian beam azimuthal charge 4 wavefront on different 5 planes. Blue curve is measurement for turbulence impacted beam shaping. Red curve is measurement with LG beams propagated through turbulence. The integer values on horizontal axis represents the position of planes according to experimental set up diagram}
    \label{Result:55}
\end{minipage}
\end{figure}

\subsection{Beam Wandering effect on Turbulence impacted beam shaping}
Further we have also plotted the graphs for beam wandering from the centroid positions for the turbulence impacts. Fig.14, Fig.15, Fig.16, Fig.17 and Fig 18 show the graphs plotted for beam wandering values for LG beams of topological charges 1 to 4 at 5 different planes respectively. The red lines in these plots show the beam wandering values for turbulence impacted LG beams of different topological charges and blue lines show the plots for the turbulence impacted laser beams converted on to LG beams of different topological charges (Beam shaping after turbulence impact) respectively. It can be seen that in Fig.14 (Plane 1)the turbulence impacted LG beams of all 4 topological charges have less beam wandering values compared to turbulence impacted laser beams shaped on to LG beams of different topological charges. At plane 2 (Fig.15) the turbulence impacted laser beams shaped on to LG beams of different topological charges have less beam wandering values except for turbulence impacted LG beams of topological charge 2. In a similar way Fig. 16(Plane 3) and 17(Plane 4) also show that the turbulence impacted laser beams shaped on to LG beams show less beam wandering effects compared to turbulence impacted LG beams except for the values obtained for the topological charge 2. Finally Fig.18(Plane 5) which is inverse Fourier Transform plane show that the turbulence impacted LG beams show less beam wandering values compared to turbulence impacted laser beams shaped on to LG beams. These findings show that the beam wandering values for laser beams shaped after the turbulence impact are more compared to the beam wandering values of turbulence impacted LG beams especially at Inverse Fourier Transform plane.  \par

\begin{figure}[H]
\centering
\begin{minipage}[b]{0.42\textwidth}
    \includegraphics[width=\textwidth]{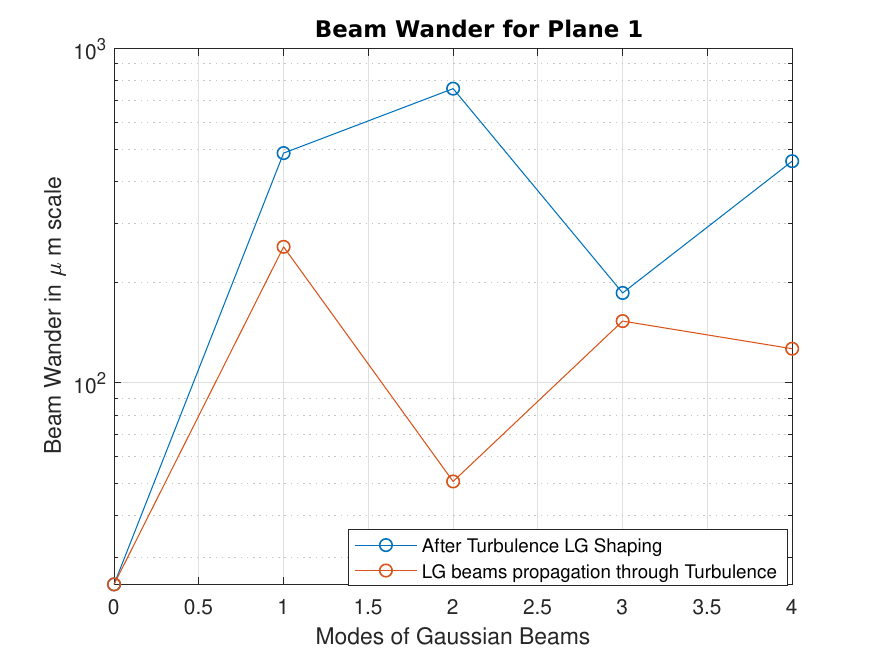}
    \caption{Beam Wander for all beams on plane 1 before application of any lens imaging system. Blue curve is measurement for after turbulence beam modulation and Red is for LG beams propagated through Turbulence. The integer values on horizontal axis represents the position of planes according to experimental set up diagram}
    \label{BW1}
\end{minipage}
\hfill
\begin{minipage}[b]{0.42\textwidth}
    \includegraphics[width=\textwidth]{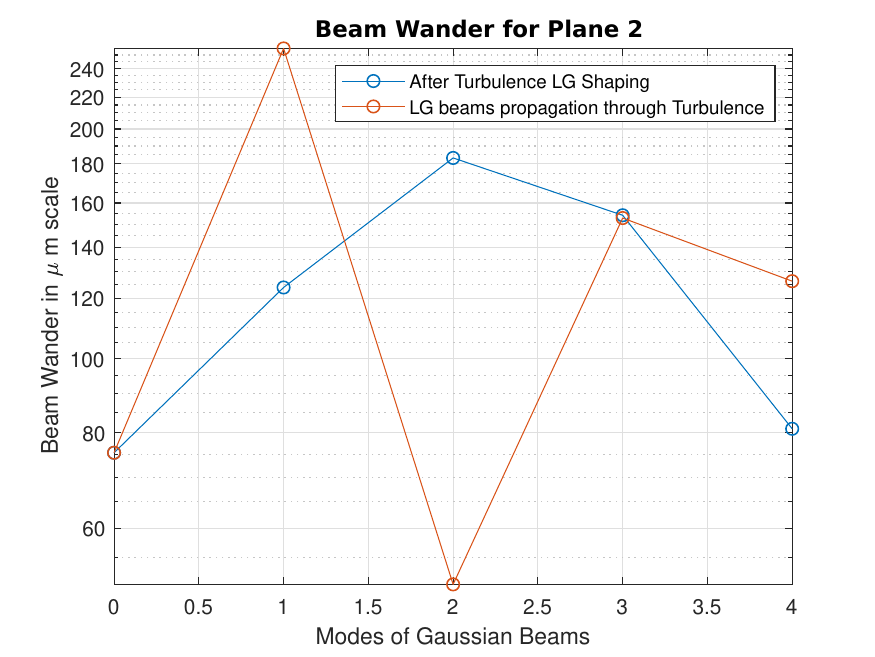}
    \caption{Beam Wander for all beams on plane 2 before application of any lens imaging system. Blue curve is measurement for after turbulence beam modulation and Red is for LG beams propagated through Turbulence. The integer values on horizontal axis represents the position of planes according to experimental set up diagram}
    \label{BW2}
\end{minipage}
\end{figure}

\begin{figure}[H]
\centering
\begin{minipage}[b]{0.42\textwidth}
    \includegraphics[width=\textwidth]{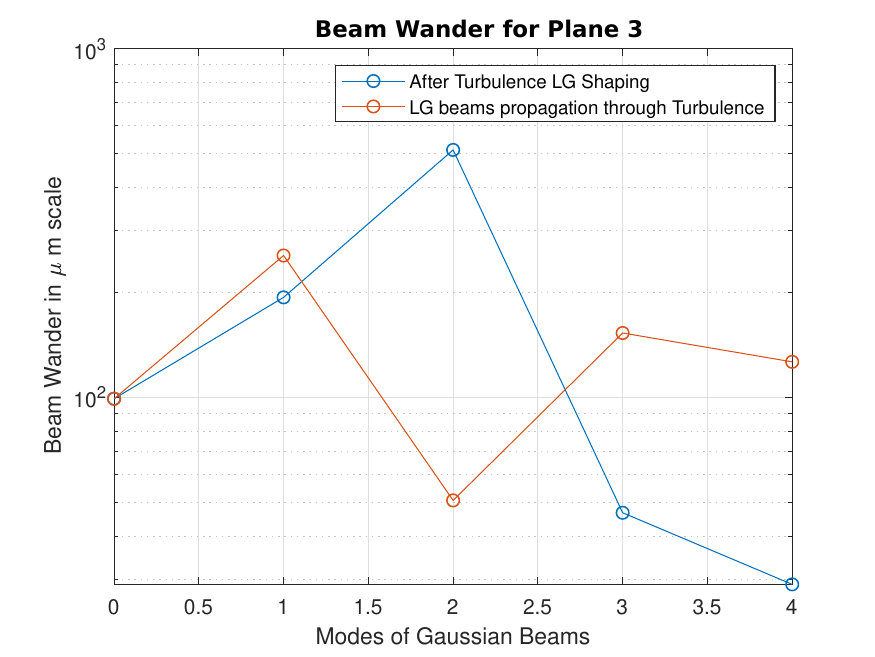}
    \caption{Beam Wander for all beams on plane 3 before application of any lens imaging system. Blue curve is measurement for after turbulence beam modulation and Red is for LG beams propagated through Turbulence. The integer values on horizontal axis represents the position of planes according to experimental set up diagram}
    \label{BW3}
\end{minipage}
\hfill
\begin{minipage}[b]{0.42\textwidth}
    \includegraphics[width=\textwidth]{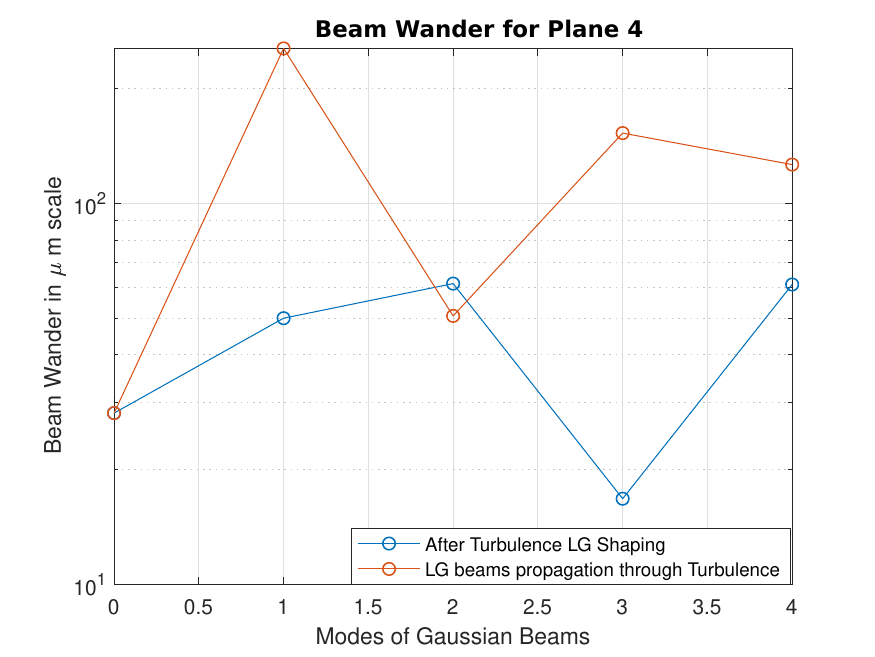}
    \caption{Beam Wander for all beams on plane 4 before application of any lens imaging system. Blue curve is measurement for after turbulence beam modulation and Red is for LG beams propagated through Turbulence. The integer values on horizontal axis represents the position of planes according to experimental set up diagram}
    \label{BW4}
\end{minipage}
\end{figure}

\begin{figure}[H]
\centering
\begin{minipage}[b]{0.42\textwidth}
    \includegraphics[width=\textwidth]{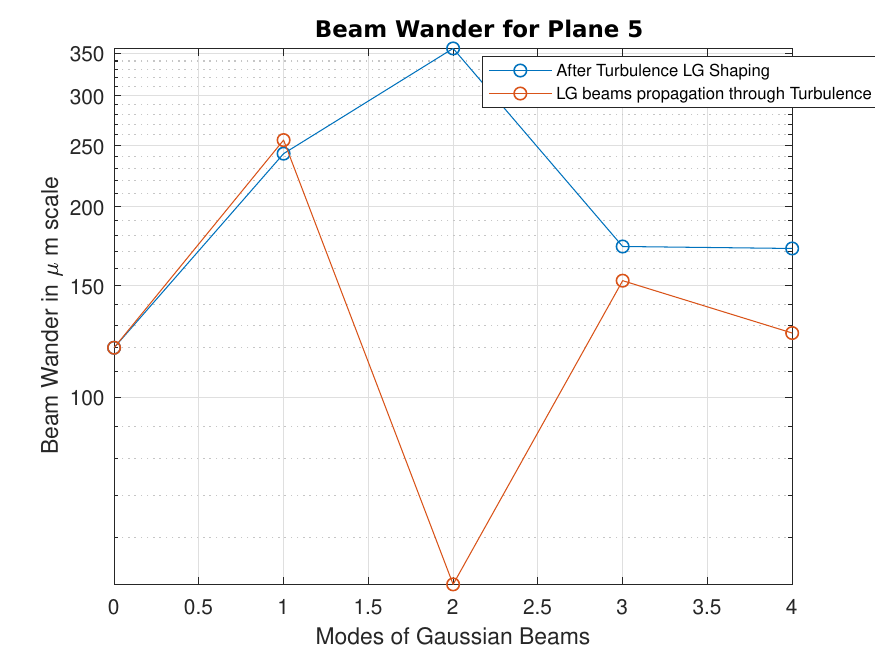}
    \caption{Beam Wander for all beams on plane 5 before application of any lens imaging system. Blue curve is measurement for after turbulence beam modulation and Red is for LG beams propagated through Turbulence. The integer values on horizontal axis represents the position of planes according to experimental set up diagram}
    \label{BW5}
\end{minipage}
\end{figure}

\section{Conclusion}

In our work to reduce the impact of turbulence on the propagating light beams through a dynamic Kolmogorov type atmospheric turbulence, instead of beam engineering before turbulence we shaped the turbulence impacted laser beams on to Laguerre Gaussian (LG) beams of different topological charges(Charge 1 to 4) and found out the scintillation index values and beam wandering values along random direction at 5 different planes along the propagation direction.  We compared the values obtained for mean scintillation index values and beam wandering values along random direction at 5 different planes for the turbulence impacted LG beams of 4 different charges with our new beam shaping technique post turbulence impact. We found that the beam shaping to different topologically charged LG beams post turbulence impacted laser beams show much less scintillation effects and more beam wandering values at plane 1(Before 4f imaging system) and plane 5(Inverse Fourier Transform plane). Thus our new method of beam shaping of turbulence impacted laser beams show that further beam shaping could be an alternative technique to conventional adaptive optics correction methods.

\section*{Funding}
Department of Science and Technology, Ministry of Science and
Technology, India (CRG/2020/003338).

\section*{Acknowledgement}
Shouvik Sadhukhan and C S Narayanamurthy Acknowledge the SERB/DST(Govt. Of India) for providing financial support via the  project grant CRG/2020/003338 to carry out this work.

\section*{Disclosure}
The authors declare no conflicts of interest.

\section*{Data availability}
All data used for this research has been provided in the manuscript itself.

\section*{Appendix}
\begin{figure}[H]
\centering
\begin{minipage}[b]{1.0\textwidth}
    \includegraphics[width=\textwidth]{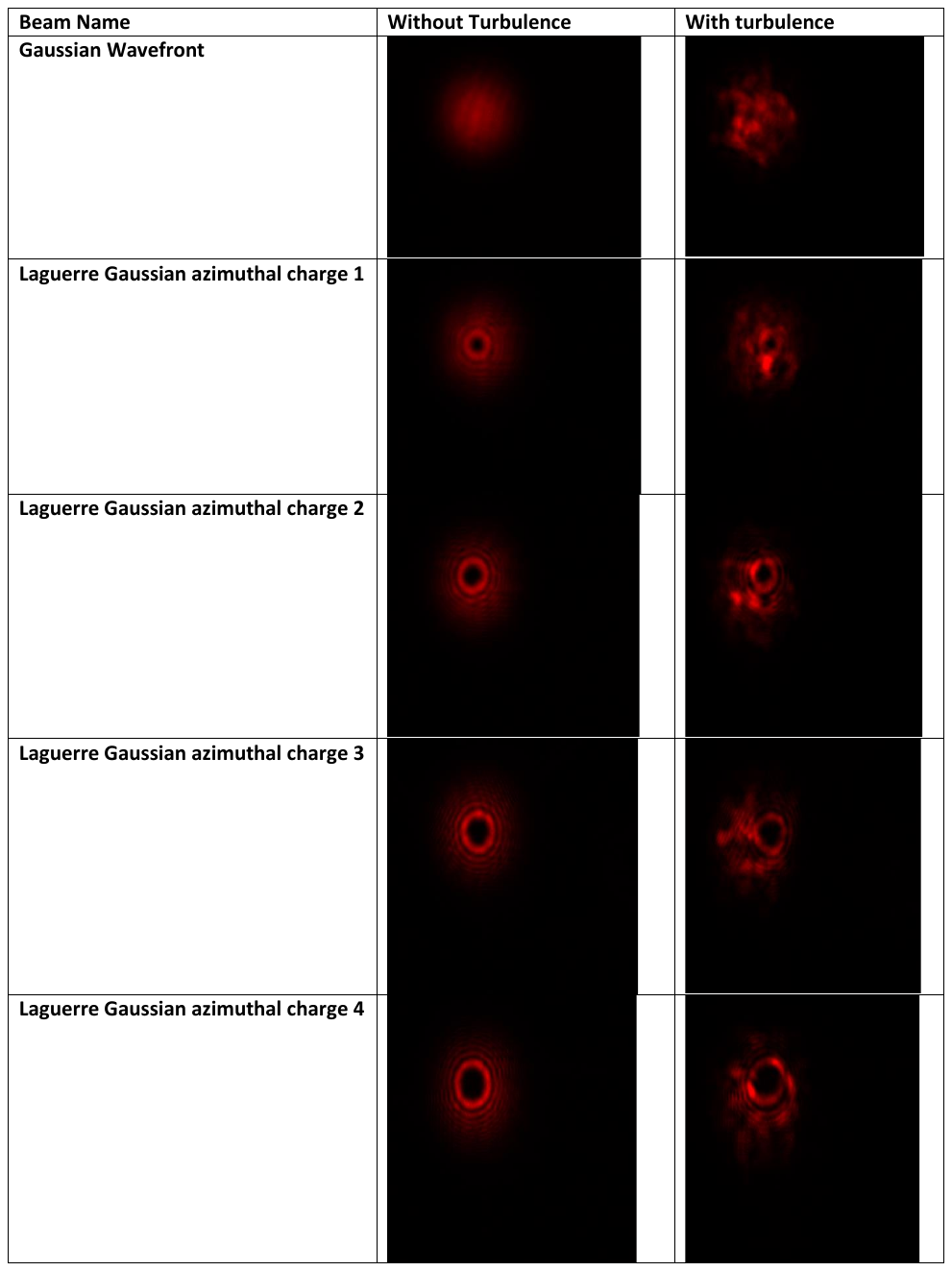}
    \caption{Five Beam (Gaussian, LG-1 to LG-4) wavefronts images without turbulence impact (Middle or 2nd Column) and turbulence impacted beam shaping (Right most column) on Plane 1}
    \label{Output:1}
\end{minipage}
\end{figure}

\begin{figure}[H]
\centering
\begin{minipage}[b]{1.0\textwidth}
    \includegraphics[width=\textwidth]{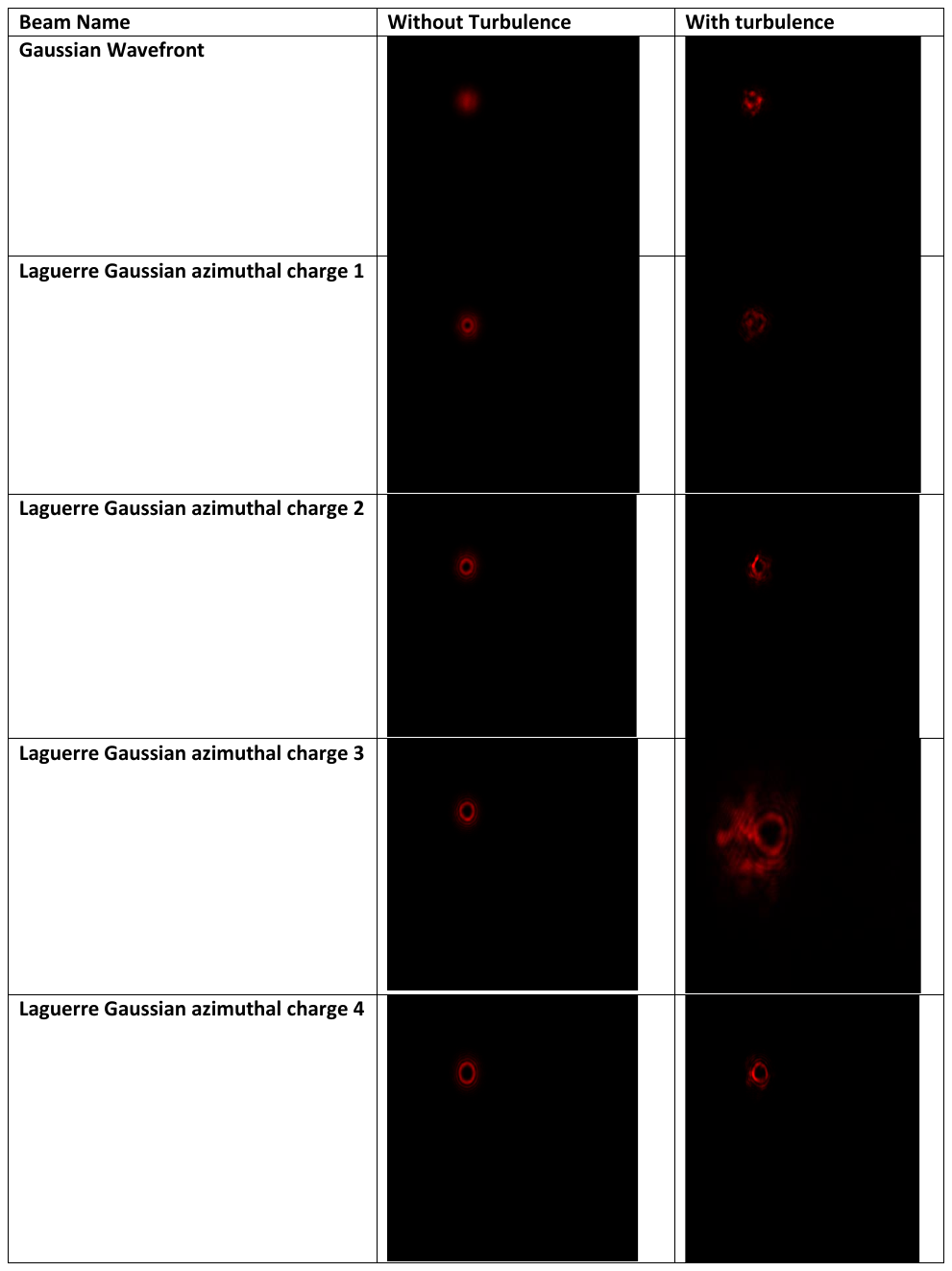}
    \caption{Five Beam (Gaussian, LG-1 to LG-4) wavefronts images without turbulence impact (Middle or 2nd Column) and turbulence impacted beam shaping (Right most column) on Plane 2}
    \label{Output:2}
\end{minipage}
\end{figure}

\begin{figure}[H]
\centering
\begin{minipage}[b]{1.0\textwidth}
    \includegraphics[width=\textwidth]{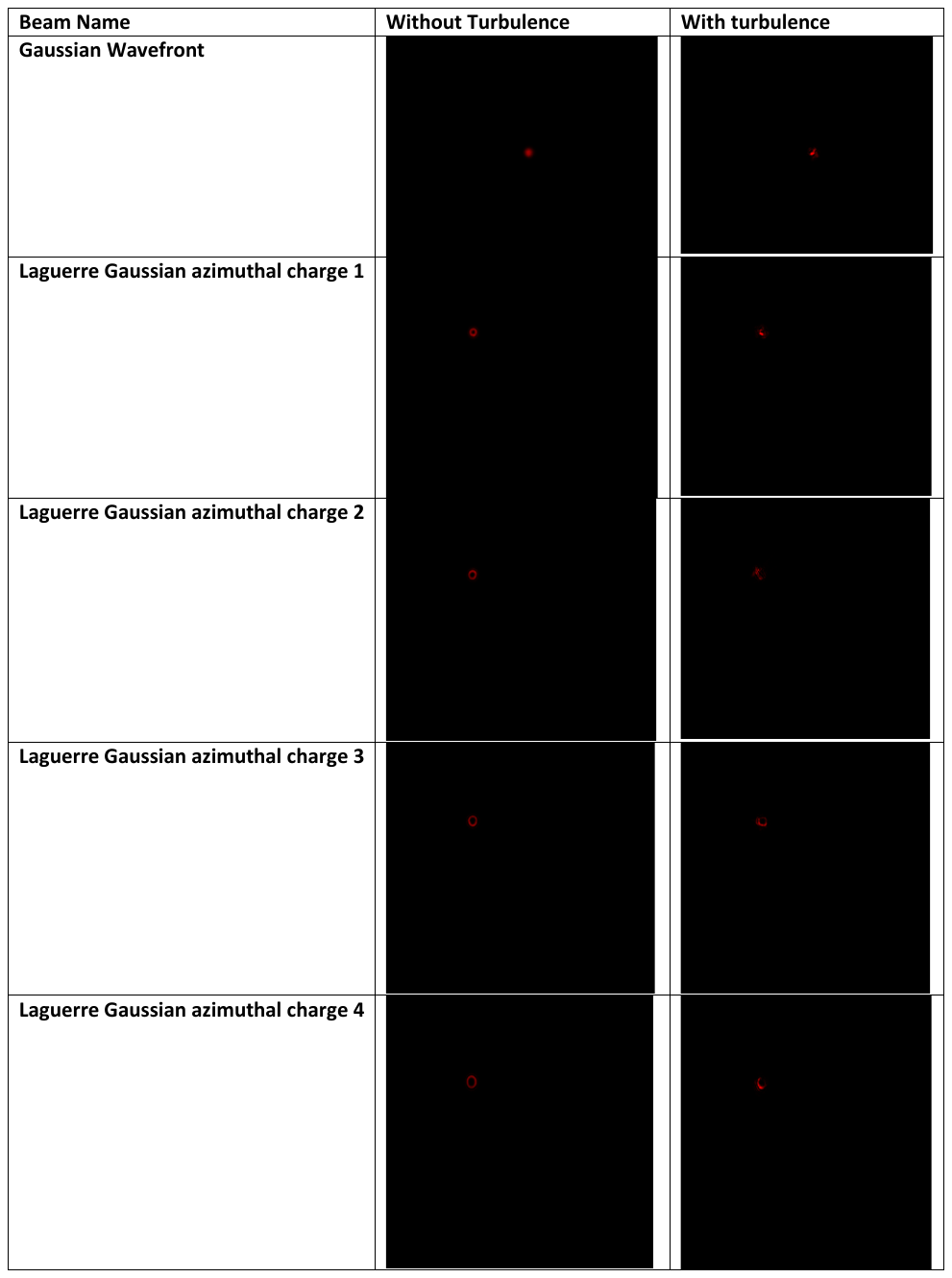}
    \caption{Five Beam (Gaussian, LG-1 to LG-4) wavefronts images without turbulence impact (Middle or 2nd Column) and turbulence impacted beam shaping (Right most column) on Plane 3}
    \label{Output:3}
\end{minipage}
\end{figure}

\begin{figure}[H]
\centering
\begin{minipage}[b]{1.0\textwidth}
    \includegraphics[width=\textwidth]{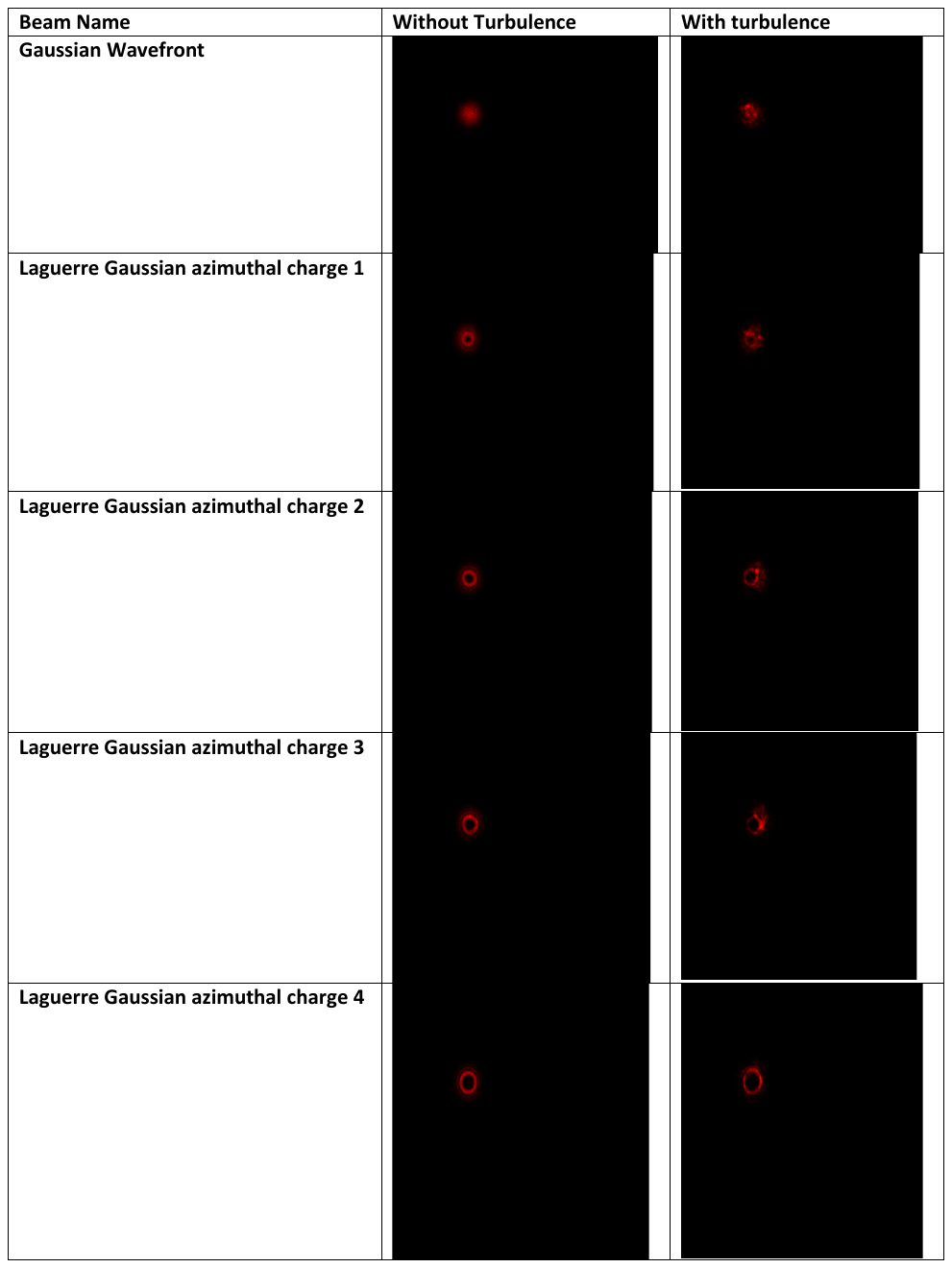}
    \caption{Five Beam (Gaussian, LG-1 to LG-4) wavefronts images without turbulence impact (Middle or 2nd Column) and turbulence impacted beam shaping (Right most column) on Plane 4}
    \label{Output:4}
\end{minipage}
\end{figure}

\begin{figure}[H]
\centering
\begin{minipage}[b]{1.0\textwidth}
    \includegraphics[width=\textwidth]{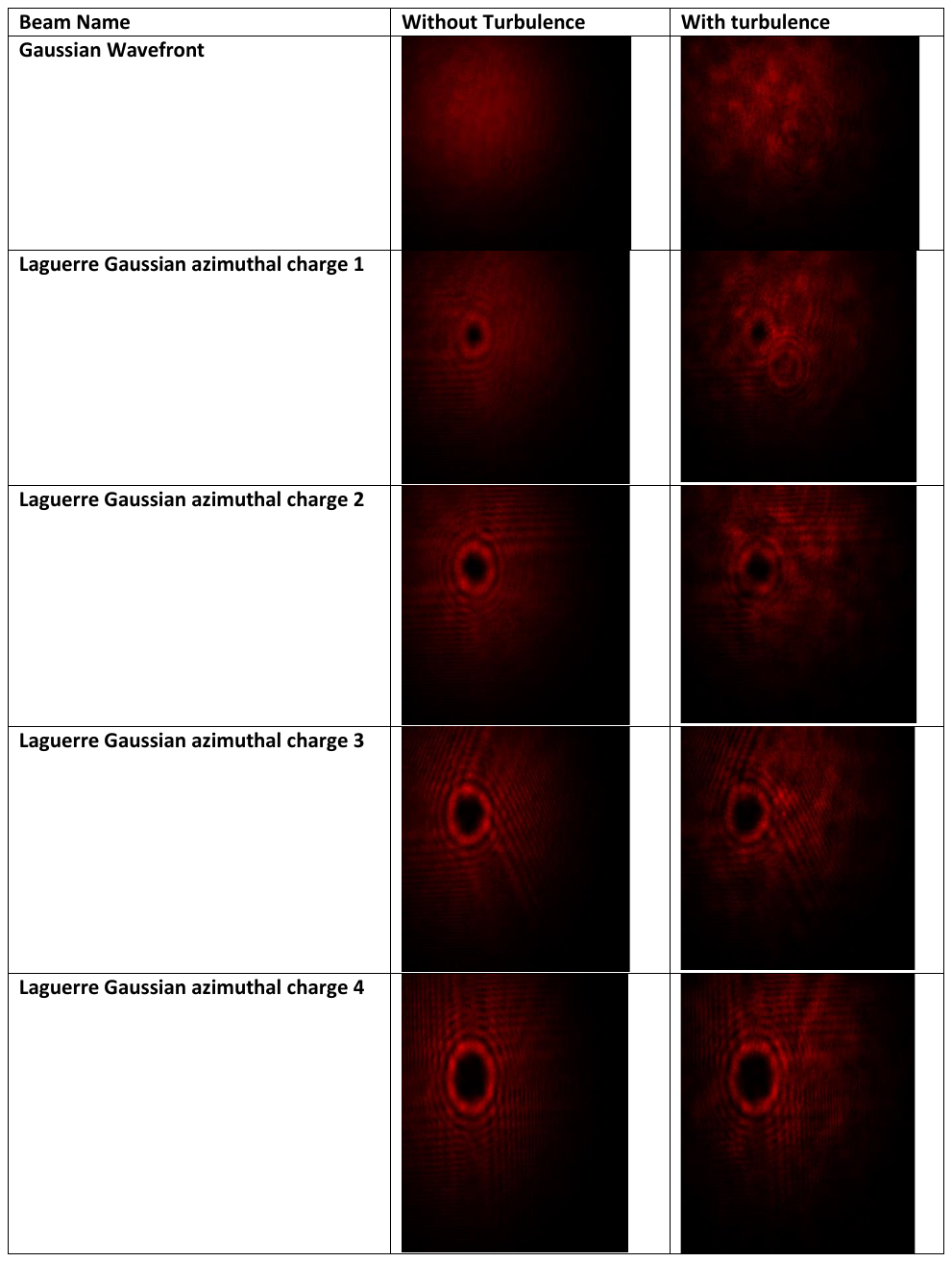}
    \caption{Five Beam (Gaussian, LG-1 to LG-4) wavefronts images without turbulence impact (Middle or 2nd Column) and turbulence impacted beam shaping (Right most column) on Plane 5}
    \label{Output:5}
\end{minipage}
\end{figure}

In this Appendix section the images obtained for Gaussian and LG beams of 4 topological charges without turbulence impact (middle column) and turbulence impacted beam shaping (right most column) at 5 different planes are shown in Fig. \ref{Output:1}(Plane 1), Fig. \ref{Output:2}, Fig.\ref{Output:3}, Fig.\ref{Output:4} and Fig.\ref{Output:5} respectively. These images clearly show the advantage of beam shaping after the turbulence impact.






\end{document}